\documentclass[twocolumn,preprintnumbers,amsmath,amssymb,showpacs]{revtex4}

\usepackage{graphicx}
\usepackage{dcolumn}
\usepackage{bm}

\newcommand{\subfig}[1]{(#1)}

\newcommand{\pic}[2]{\begin{picture}(0,#1)(0,0)#2\end{picture}}
\newcommand{\putpic}[2]{\put(#1){\begin{picture}(0,0)(0,0)#2\end{picture}}}

\begin{document}
\title{Geometrical families of mechanically stable granular packings}

\author{Guo-Jie Gao$^1$}
\author{Jerzy B{\l}awzdziewicz$^{1,2}$}
\author{Corey S. O'Hern$^{1,2}$}
\affiliation{$^1$~Department of Mechanical Engineering, Yale University,
New Haven, CT 06520-8284.\\
$^2$~Department of Physics, Yale University, New Haven, CT 06520-8120.\\
}
\date{\today}

\begin{abstract}
We enumerate and classify nearly all of the possible mechanically
stable (MS) packings of bidipserse mixtures of frictionless disks in
small sheared systems.  We find that MS packings form continuous
{\it geometrical families}, where each family is defined by its particular
network of particle contacts.  We also monitor the dynamics of MS
packings along geometrical families by applying quasistatic simple
shear strain at zero pressure.  For small numbers of particles ($N <
16$), we find that the dynamics is deterministic and highly
contracting.  That is, if the system is initialized in a MS packing at
a given shear strain, it will quickly lock into a periodic orbit at
subsequent shear strain, and therefore sample only a very small
fraction of the possible MS packings in steady state.  In studies with
$N>16$, we observe an increase in the period and random splittings of
the trajectories caused by bifurcations in configuration space.  We
argue that the ratio of the splitting and contraction rates in large
systems will determine the distribution of MS-packing geometrical
families visited in steady-state. This work is part of our long-term
research program to develop a master-equation formalism to describe
macroscopic slowly driven granular systems in terms of collections of
small subsystems.
\end{abstract}

\maketitle

\section{Introduction}
\label{introduction}

Dry granular materials are collections of discrete, macroscopic
particles that interact via dissipative and purely repulsive
interactions, which are nonzero when particles are in contact and
vanish otherwise.  Granular systems range from model systems composed
of glass beads to pharmaceutical powders, to soils and geological
materials.  

A distinguishing feature of granular materials is that
they are {\it athermal}.  Since individual grains are large, thermal
energy at room temperature $T$ is unable to displace individual
grains.  Thus, without external driving, granular materials are static
and remain trapped in a single mechanically stable (MS) grain packing
with force and torque balance on each grain.
In contrast, when external forces are applied to granular materials,
these systems flow, which gives rise to grain rearrangements,
fluctuations in physical quantities like shear stress and pressure,
and the ability to explore configuration space.  

There are many driving mechanisms that generate dense flows in
granular media---for example, oscillatory
\cite{toiya,dauchot,pouliquen,behringer2} and continuous
shear~\cite{nicolas,marty}, horizontal \cite{ristow} and vertical
vibration \cite{nowak,philippe}, and gravity-driven flows \cite{choi}.  The
fact that driven granular systems can achieve steady-states, explore
configuration space, and experience fluctuations as in thermal
systems, has prompted a number of groups to describe these flows using
concepts borrowed from equilibrium statistical mechanics (such as
effective temperature) \cite{makse2,ono,ohern,makse3,makse4,henkes}.

However, before a statistical mechanical treatment can be rigorously
applied to dense granular flows, fundamental questions about the
nature of configuration space should be addressed.  In particular, one
needs to determine how dense granular systems sample configuration
space: Is it uniformly sampled or are some states visited much more
frequently than others?  How is the sampling of configuration space
affected by the strength and type of driving and dissipation
mechanisms?  In this work and a series of recent papers
\cite{xu2,gao,phil_mag,marks}, we address these questions with the
goal of developing a comprehensive physical picture for static and
slowly driven granular matter.

In our previous studies, we focused on the statistical properties of
static frictionless disk packings generated by slow compression
without gravity \cite{xu2,gao,phil_mag} or by gravitational deposition
\cite{marks}.  We have determined that the probability distribution
for mechanically stable packings is strongly peaked around the value
typically quoted for the random-close packing (RCP) volume fraction,
and explained why RCP is obtained by many compaction protocols.  We
have also found that the MS-packing probabilities are highly
non-uniform, contrary to the Edwards' equal-probability assumption
\cite{edwards} that is frequently used in thermodynamic descriptions
of granular matter \cite{makse3}.  In addition, we have found that the
probabilities become more non-uniform with increasing system
size~\cite{gao}.

In the present article, we further explore the statistics of granular
microstates and its relevance for static and dynamic properties of
granular materials.  We focus on slowly sheared systems at fixed zero
pressure, where the evolution can be approximated as a sequence of
transitions between MS packings.  One of our novel results is that
slowly sheared MS packings occur as continuous {\it geometrical
families} defined by the network of particle contacts.  Moreover,
these geometrical families are not uniformly sampled during
quasistatic shear flow. We focus on small systems, so that we are able
enumerate nearly all MS packings and obtain accurate packing
probabilities as a function of shear strain.  Since our results
indicate the need for developing an alternative approach to the
quasi-thermodynamic descriptions based on the Edwards' ideas, we
advocate here a new master-equation framework for understanding 
dense granular flows.

This paper is organized as follows.  In Sec.~\ref{Motivation}, we
provide motivation for our investigations by introducing a simple
phenomenological master-equation model that reproduces qualitatively
some of the key features of slowly-driven granular systems.  In
Secs.~\ref{system} and ~\ref{enumeration}, we introduce our model
system, 2D bidisperse mixtures of frictionless, purely repulsive, soft
grains, and describe the simulation method that we employ to generate
mechanically stable packings. Here, we also clearly define the set of
distinct mechanically stable (MS) packings in terms of the eigenvalues
of the dynamical matrix and discuss their symmetries.  In
Sec.~\ref{Quasistatic evolution}, we outline our method to study
quasistatic simple shear flow of frictionless disks in small 2D
systems at zero pressure.  In Secs.~\ref{families}, we describe the
results of the quasistatic shear simulations, with a particular
emphasis on enumerating geometrical families of MS packings and
determining how they are sampled during quasistatic dynamics. In
Sec.~\ref{random splitting} provide an outlook for further work 
on larger system sizes. The
main conclusions of our studies and their relation to our long-term
research program are discussed in Sec.~\ref{conclusions}. In
Appendices~\ref{Numerical details}, \ref{matrix}, and \ref{Appendix:
polarizations}, we provide details of the numerical simulations, the
calculation of the dynamical matrix for the repulsive linear spring
potential used in these studies, and the method used to distinguish
`polarizations' of MS packings ({\it i.e.}, MS packings that differ
only by reflection or rotation).

\section{Motivation}
\label{Motivation}

\begin{figure}
\includegraphics[width=0.4\textwidth]{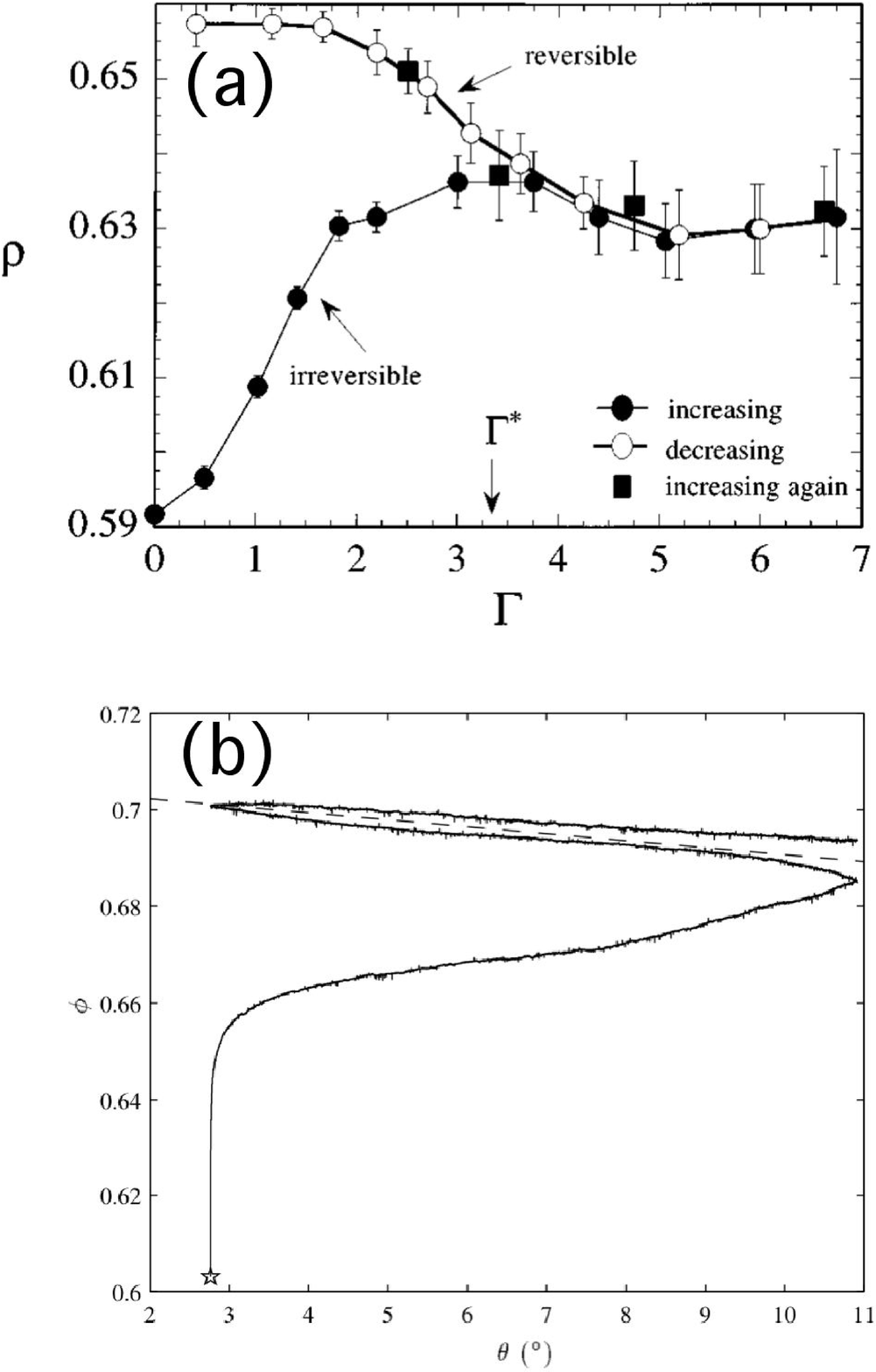}

\caption{ \subfig{a} Packing density $\rho$ for
  mm glass beads as a function of tapping intensity $\Gamma$
  (normalized by $g$).  The system is initialized in a dilute state at
  $\rho=0.59$ and then subjected to vibrations of increasing
  intensity.  At each $\Gamma$, the system is tapped until it achieves
  a steady-state $\rho$.  After reaching $\Gamma=7$, the tapping
  intensity is decreased until $\Gamma<1$, and then increased again.
  Data is reprinted from Ref.~\cite{nowak}. Copyright 1998, The
  American Physical Society.  \subfig{b} Packing fraction $\phi$ for
  mm glass beads as a function of shear angle $\theta$ during cyclic
  shear.  $\theta$ is increased linearly in time from $2.7^{\circ}$ to
  $10.7^{\circ}$, then decreased to $2.7^{\circ}$, and finally increased
  again to $10.7^{\circ}$.  Data is reprinted from
  Ref.~\cite{nicolas}. Copyright 2000, Springer-Verlag.}
\label{Reversible and irreversible branches}
\end{figure}
\begin{figure}
\pic{130}{


\putpic{-120,-10}{
\put(15,15){
\includegraphics*[width=0.40\textwidth,trim=0 0 0 0]
		    {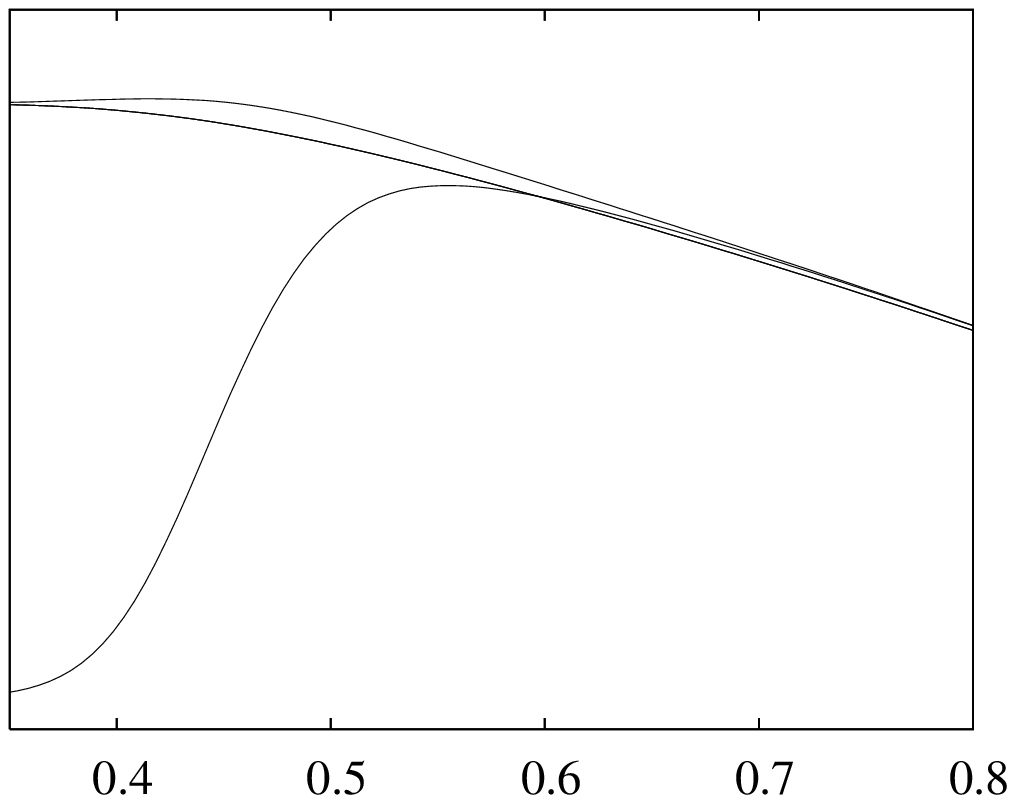}}

\put(140,5){\small $\lambda$}
\put(30,90){\rotatebox{0}{\small $\phi$}}

\put(67,60){\small\it I}
\put(65,120){\small\it R}

\putpic{90,40}{
\put(0,0){
\includegraphics*[width=0.18\textwidth,trim=0 0 0 0]
		    {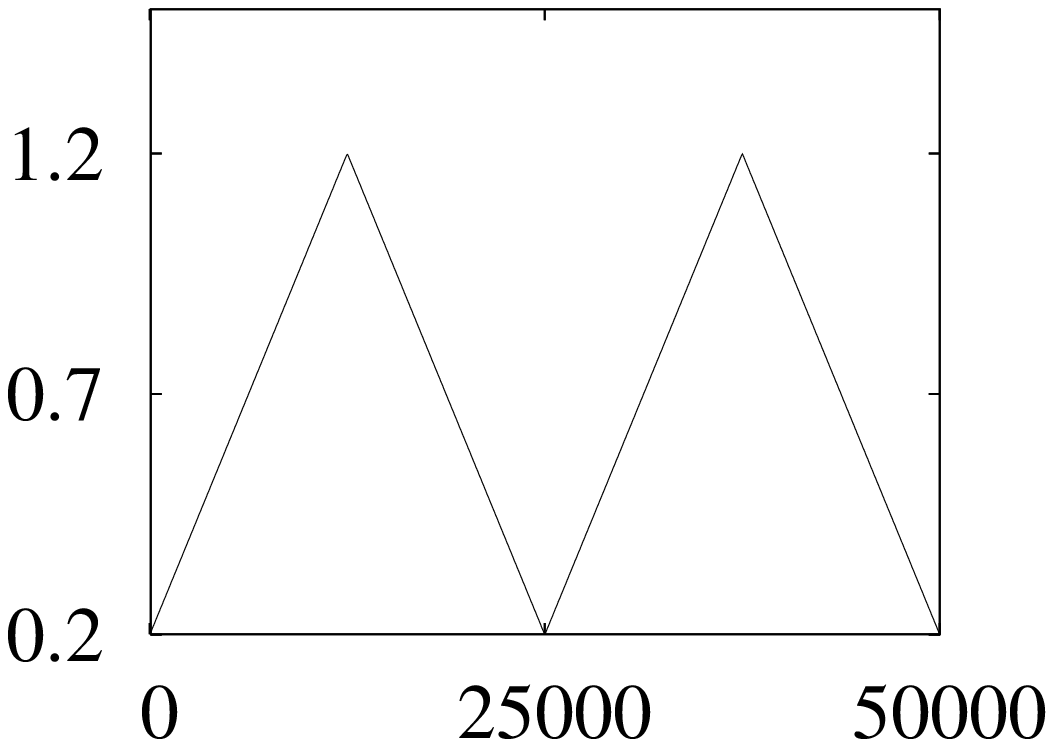}}
\put(55,-3){\footnotesize $t$}
\put(5,28){\footnotesize $\lambda$}

}

}


}
\caption{\label{quasistatic evolution} Model for quasistatic evolution
of slowly driven granular materials.  Evolution of a system undergoing
a long sequence of quasistatic excitations (taps).  The excitation
amplitude $\lambda$ is periodically ramped up and down in time $t$, as
shown in the inset.  The system initially evolves along the
irreversible branch $I$, but at subsequent periods of ramping it moves
along the reversible branch $R$ (with slight hysteresis).  The results
were obtained by solving a set of master equations \eqref{Master
equation} with model transition rates $W_{kl}$ (between states $k$ and
$l$) adjusted to qualitatively reproduce the packing fraction versus
tapping intensity obtained in experiments shown in
Fig.~\ref{Reversible and irreversible branches} \subfig{a}.}
\vspace{-0.05in}
\end{figure}

Granular materials are athermal---they are unable to thermally
fluctuate and sample phase space.  However, if a grain packing is
perturbed by external forces, it can move through a series of
configurations.  The set of states populated by a granular system
during a series of discrete vertical taps was characterized in
Ref.~\cite{nowak}. As shown in Fig.~\ref{Reversible and irreversible
branches}\subfig{a}, an initially loose packing is compacted by
tapping first gently, and then with greater intensity. At sufficiently
large tapping intensities, it is no longer possible to further compact
the system. However, when the tapping intensity is reduced, the
packing fraction increases, rather than returning along the original
packing fraction trajectory.  This new curve (packing fraction
vs. tapping intensity) obtained by successively decreasing and then
increasing the tapping intensity in small steps is nearly reversible.
A similar phenomenon has been found in granular media undergoing
cyclic shear \cite{nicolas}, as shown in Fig.~\ref{Reversible and
irreversible branches}\subfig{b}.

These experiments, which show that slowly driven granular systems
appear to explore a well-defined set of states reversibly, have
prompted a number of theoretical studies aimed at describing
compacting granular systems using quasi-thermodynamic approaches based
on the Edwards' ensemble ({\it i.e.} the assumption of equally probable
microstates)~\cite{depken,tarjus,barrat2,a_barrat,fierro}.  However, in
previous work \cite{xu2,gao,phil_mag,marks}, we have shown explicitly
for small systems that the probability distribution for mechanically
stable packings is highly non-uniform.  Moreover, we have
demonstrated that this feature is not sensitive to the
packing-generation protocol and becomes more pronounced as the system
size increases. Thus, we argue that the Edwards' equal-probability
assumption is not valid and alternate theoretical approaches for
slowly driven granular systems must be developed.

Although several alternatives have been put forward
\cite{henkes,henkes2,dauchot2}, we advocate here for a master-equation approach
for the following reasons. First, quasistatic evolution of slowly
driven granular systems can be approximated as a sequence of
transitions between MS packings.  Second, since the system undergoes particle
rearrangements as it transitions between MS packings, it
retains little memory from one MS packing to the next, and successive
MS packings are nearly statistically independent.  Thus, slowly driven
granular systems can be approximated as a Markov chain of independent
transitions between MS packings and described using a master-equation
approach.

We have shown that the form of the microstate probability distribution
can be qualitatively reproduced by combining probabilities for small
subsystems.  Thus, we also advocate a novel `bottom-up'
approach in which large granular systems are described as collections
of nearly independent subsystems.  We view this work on small
quasistatically sheared MS packings as laying the groundwork for
future studies that will apply the master-equation approach to
quantitatively describe the statistical properties of dense granular
flows.

\newcommand{\Nsubsyst}{{\overline N}}
\newcommand{\Nmicrostates}{m}

\subsection{Model}
\label{theory_model}

To illustrate the importance of the above-mentioned features in
capturing the irreversible and reversible branches in
Fig.~\ref{Reversible and irreversible branches}, we construct a simple
model in which a granular system is represented as a collection of
$\Nsubsyst$ statistically independent small subdomains.  Each
subdomain $j=1,\ldots,\Nsubsyst$ can reside in one of several
microstates $k_j=1,\ldots,\Nmicrostates$.  The volume of the subdomain
$j$ in state $k_j$ is $V_1(j,k_j)$, and these subvolumes are assumed
to be additive
\begin{equation}
\label{additive subvolumes}
V(\Lambda)=\sum_{j=1}^\Nsubsyst V_1(j,k_j),
\end{equation}
where $V(\Lambda)$ is the volume of the whole system in the state
$\Lambda=(k_1,\ldots,k_\Nsubsyst)$.  Since the subdomains are
assumed to be statistically independent, the joint probability
distribution for the whole system is
\begin{equation}
\label{joint probability}
P(\Lambda;t)=\prod_{j=1}^\Nsubsyst P_j(k_j;t),
\end{equation}
where $P_j(k_j;t)$ is the probability distribution for the microstates of
subdomain $j$ at time $t$.  

The evolution of subdomains in a system driven by an applied force of
strength $\lambda$ is described by $\Nsubsyst$
independent master equations
\begin{eqnarray}
\label{Master equation}
P_j(k;t_{i+1})=P_j(k;t_{i})+\sum_{k'=1}^\Nmicrostates
  [W_{kk'}(\lambda)P_j(k';t_i)-&&\nonumber\\
W_{k'k}(\lambda)P_j(k;t_i)],&&
\end{eqnarray}
where $W_{kk'}$ is the transition probability from state $k'$ to state
$k$.

To qualitatively reproduce the irreversible and reversible branches of
states when the magnitude of the external forcing $\lambda$ is ramped
up and down, we use simple assumptions regarding the volume
distribution of individual subsystems and the transition
probabilities.  We assume that the volumes of individual subsystems
are given by the expression
\begin{equation}
\label{volume of subsystem}
V_1=r(V_0+A\mathrm{e}^{-\kappa_v x_i}),
\end{equation}
where $V_0$, $A$, and $\kappa_v$ are constants, $x_i=(i-1)/(k-1)$, and
$r$ is a random number. Note that states with higher indexes possess 
smaller volumes. The transition rates are modeled by the expression
\begin{equation}
\label{transition rate}
W_{ij}=C b(x_{ij}) r_w \textrm{e}^{-[x_{ij}/\sigma_0\lambda(t)]^2},
\end{equation}
where $C$ is the normalization constant, $b(x_{ij})$ is a bias
function 
\begin{equation}
\label{bias function}
b(x)=\left\{
  \begin{array}{ll}
    b_{\textrm{min}}+b_0(1-2\lambda)&\qquad x>0,
\\
    1&\qquad x<0
  \end{array}
\right.
\end{equation}
with constants $b_{\rm min}$ and $b_0$, which introduces asymmetry
between up and down jumps, $\sigma_0$ determines the width of the
Gaussian distribution, and $r_w$ is a random number.  As shown in
Fig.~\ref{quasistatic evolution}, such a simple master-equation
approach is able to qualitatively reproduce features (evolution of the
system to a reversible branch of the packing fraction) found in
vertical tapping and cyclic shear studies of granular materials.  The
success of this simple model emphasizes two key points: 1. Only
minimal constraints on transition probabilities are required (not
thermodynamics) to reach steady-state and 2. An assumption of weakly
interacting small subsystems may be able to explain macroscopic
phenomena in slowly driven granular media.

These results suggest a new approach to describe the quasistatic
evolution of granular systems: a Markov process characterized by
transfer rates between MS packing microstates.  To further develop
this approach so that it can yield quantitative predictions, one must
determine (a) the types of microstates that occur in static granular
systems and (b) transition probabilities between these microstates
when the system is slowly driven.  To provide the necessary input for
constructing quantitative descriptions, in this paper we study the
structure of configuration space and transitions between microstates
in small 2D granular systems undergoing quasistatic shear strain at 
fixed zero pressure.

\begin{figure}
\scalebox{0.15}{\includegraphics{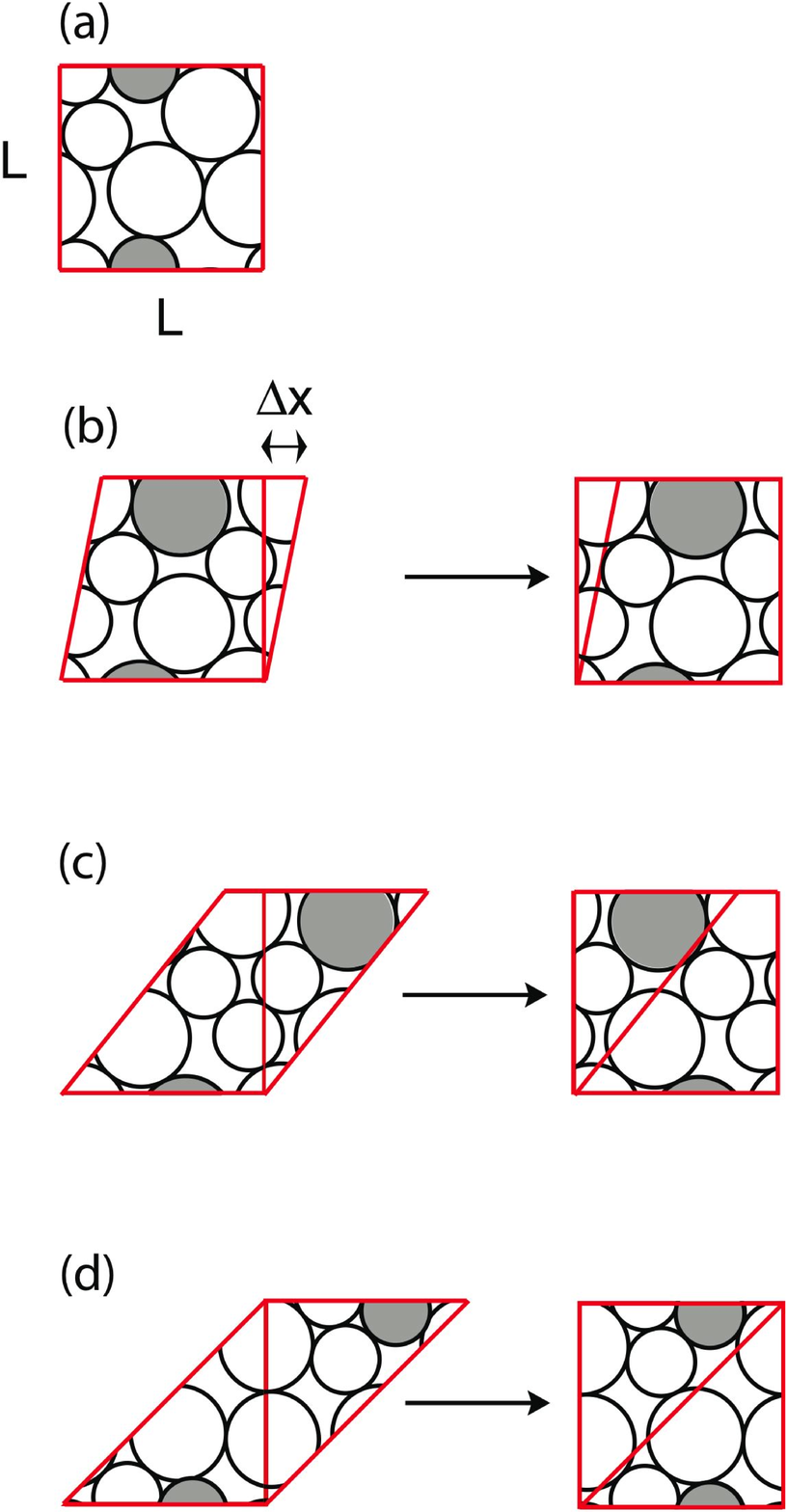}}%
\vspace{-0.18in}
\caption{\label{fig:units_strains} Schematic of shear-periodic
boundary conditions. (a) MS packing with $N=6$ particles confined to a
$L \times L$ box with shear strain $\gamma = \Delta x/L = 0$.  (b) MS
packing in a unit cell with $\gamma = 0.2$.  Note that at arbitrary
$\gamma$, a given particle in the primary cell is not directly above
(or below) its image. (c) MS packing at $\gamma = 0.8$, which shows
that configurations at $\gamma$ and $1-\gamma$ are related by an
inversion about the vertical axis. (d) MS packing in (a) at
$\gamma=1$.  At unit shear strain, shear-periodic boundary conditions are
identical to standard periodic boundary conditions \cite{allen}.  Thus
shear-periodic boundary conditions have unit period. In all panels,
the shaded particles indicate a given particle in the primary cell and its 
image in a neighboring cell.}
\vspace{-0.22in}
\end{figure}

\begin{figure}
\scalebox{0.3}{\includegraphics{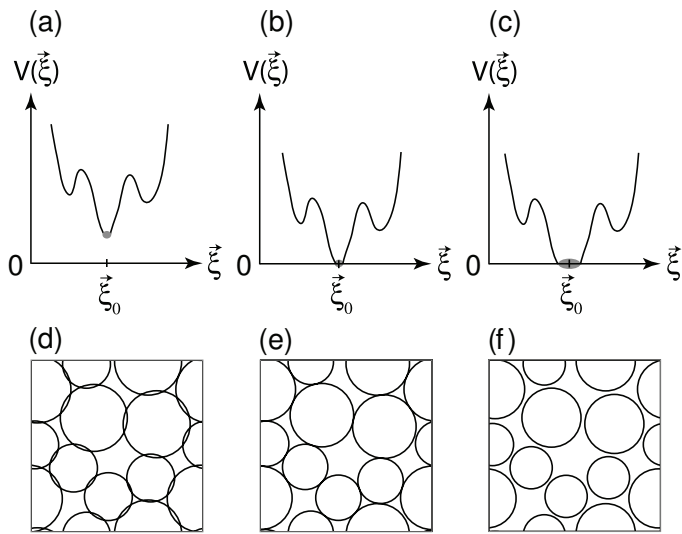}}%
\vspace{-0.0in}
\caption{\label{fig:energy} Schematic of the compression/decompression
protocol to create MS packings. In panels (a)-(c), we show a schematic of the
energy landscape $V({\vec \xi})$ in the vicinity of the 
static granular packings (at point ${\vec \xi}_0$ in configuration
space) in panels (d)-(f).  If the system exists in a non-overlapped
configuration (panel (f)) with gaps between particles and zero
energy per particle $(V=0)$, it will be compressed followed
by energy minimization.  If the system exists in an overlapped
configuration (panel (d)) at a local energy minimum with
$V>0$, it will be decompressed followed by energy minimization.  When
the system switches between the cases displayed in panels (d) and (f),
the size of the compression/decompression increment is decreased.  The
process stops when the system exists in a static granular packing at a
local energy minimum that is infinitesimally above zero.
This schematic is shown for shear strain $\gamma=0$, but a similar
process occurs for each $\gamma$.}
\vspace{-0.22in}
\end{figure}

\begin{figure}
\scalebox{0.4}{\includegraphics{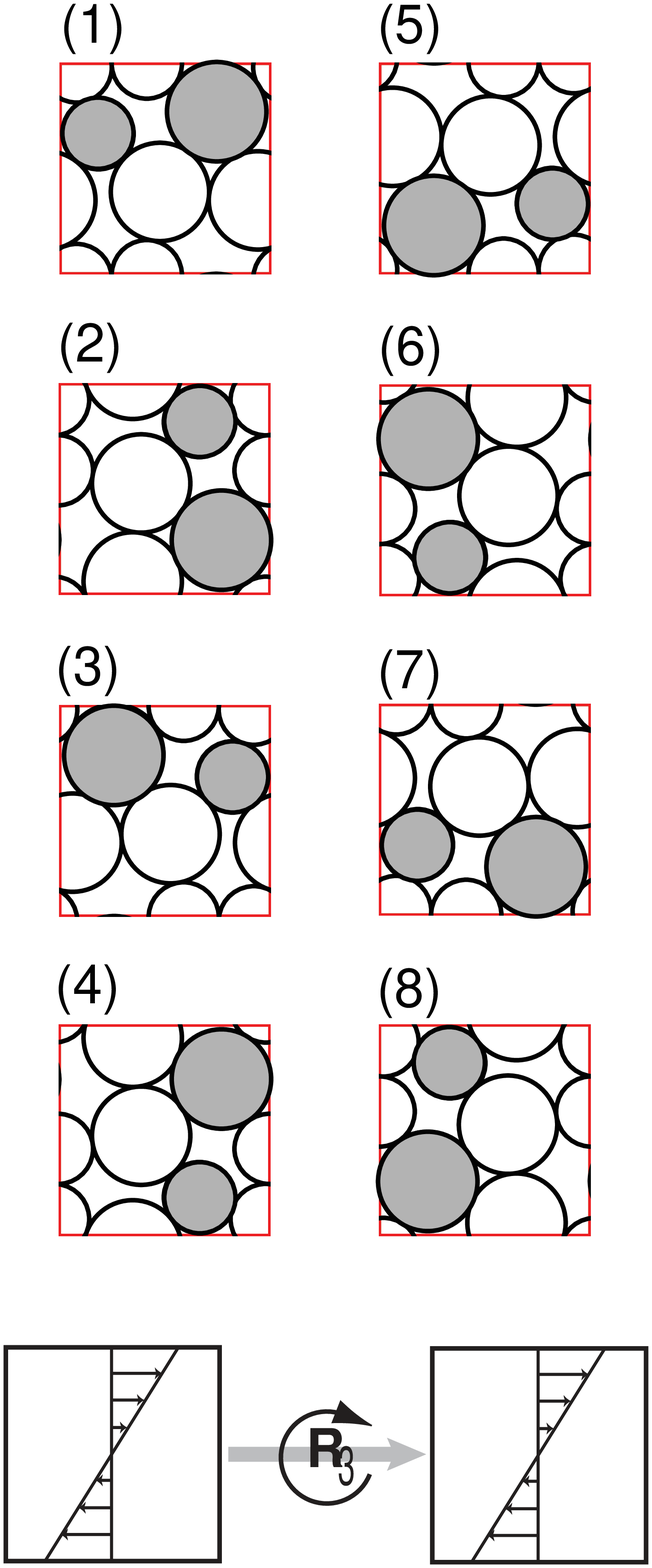}}%
\caption{\label{fig:polarization_noshear} (1) A typical MS packing for
$N=6$ particles at integer shear strain. The configurations in panels
(2)-(8) are obtained by applying the seven possible reflections and
rotations in two spatial dimensions consistent with periodic boundary
conditions in an undeformed square cell.  (See Appendix~\ref{Appendix:
polarizations}.)  Configurations in opposing columns are related by a
rotation ${\bf R}_3$ by $\pi$ about an axis coming out of the page.
(bottom) This schematic shows that configurations undergoing simple
planar shear in 2D are invariant with respect to ${\bf R}_3$.}
\vspace{-0.22in}
\end{figure}

\section{Small periodic granular packings in simple shear}
\label{system}

\subsection{Bidisperse frictionless disks}

We consider 2D systems of soft, frictionless disks interacting via the pairwise
additive purely repulsive linear spring potential
\begin{equation}
\label{total energy}
V({\vec \xi})=\sum_{i<j=1}^N V_2(r_{ij}),
\end{equation}
\begin{equation}
\label{spring potential}
V_2(r_{ij})
=\frac{\epsilon}{\alpha}\left(1-r_{ij}/\sigma_{ij}\right)^\alpha
\Theta\left(\sigma_{ij}/r_{ij}-1\right),
\end{equation}
where $\alpha=2$, $\vec \xi=(\vec r_1,\ldots,\vec r_N$) denotes the system
configuration, $\vec r_i$ is the position of the center of disk $i$,
$r_{ij}=|\vec r_i-\vec r_j|$ is the center-to-center separation
between disks $i$ and $j$, and the sum is over distinct particle
pairs.  The strength of the spring potential \eqref{spring potential}
is characterized by the energy scale $\epsilon$, and the range by the
average particle diameter $\sigma_{ij}=\left(\sigma_i + \sigma_j
\right)/2$.  The Heaviside step function $\Theta(x)$ turns off the
interaction potential when the particle separation is larger than
$\sigma_{ij}$.

All numerical simulations described in this paper were performed for
50-50 (by number) binary mixtures of large and small particles with
diameter ratio $1.4$. In such bidisperse mixtures, shear-induced
crystallization is inhibited \cite{xu}; thus the system is well suited
for investigations of quasistatic evolution of disordered granular
systems.  We focused on small systems with the number of particles in
the range $N=4$ to $20$.

To mimic the behavior of granular packings, we consider MS disk
configurations at infinitesimal pressure and particle overlaps.  As
shown in our recent experimental and numerical study, statistical
properties of disks interacting via the repulsive linear spring
potential \eqref{spring potential} closely match properties of plastic
and steel disks in a system where frictional forces have been relaxed
using high-frequency, low-amplitude vibrations \cite{marks}.

\subsection{Shear-periodic boundary conditions}

In our simulation studies, the particles are confined to a $L\times L$ periodic
box, as illustrated in Fig.\ \ref{fig:units_strains}.  The unit cell
can either be a square [cf.\ Fig.\ \ref{fig:units_strains}\subfig{a}],
or it can be deformed in the $x$ direction [cf., Figs.\
\ref{fig:units_strains}\subfig{b} and \subfig{c}].  These
shear-periodic boundary conditions allow us to generate an ensemble of
{\it anisotropic} granular packings as a function of the shear strain
$\gamma = \Delta x/L$, where $\Delta x$ is the horizontal shift of the
top image cells relative to the bottom image cells.  Moreover,
simulations of systems with gradually changing strain \cite{allen,evans}
enable us to study quasistatic evolution of a granular packing under
shear.

Note that shear-periodic boundary conditions are identical at $\gamma$
and $1-\gamma$ as shown in Fig.~\ref{fig:units_strains}. Also, when
reflection symmetry is taken into account, it is clear that we only
need to consider shear strains in the range $\gamma = [0,0.5]$ to
generate static packings at arbitrary shear strains.  However, in the
case of continuous quasistatic shear flow, evolution of the system
over multiple shear strain units must be investigated to capture the
full dynamics.

\begin{table}
\caption{\label{tab:table5} Number of distinct MS packings at integer
shear strains when we treat all polarizations as the same $N_s$ and
when we distinguish among different polarizations $N_s^p$ as a
function of system size $N$. The third column gives the ratio
$N_s^p/N_s$.  Data for $N_s$ is obtained from Ref.~\cite{gao}.}
\begin{ruledtabular}
\begin{tabular}{rrrr}
$N$ & $N_s$ & $N_s^p$ & $N_s^p/N_s$ \\
\tableline
$4$ & $3$ & $6$ & $2.00$\\ 
$6$ & $20$ & $68$ & $3.40$ \\
$8$ & $165$ & $612$ & $3.71$ \\
$10$ & $1618$ & $6378$ & $3.94$ \\
$12$ & $23460$ & $91860$ & $3.92$ \\
\end{tabular}
\end{ruledtabular}
\end{table}

\begin{figure}
\scalebox{0.38}{\includegraphics{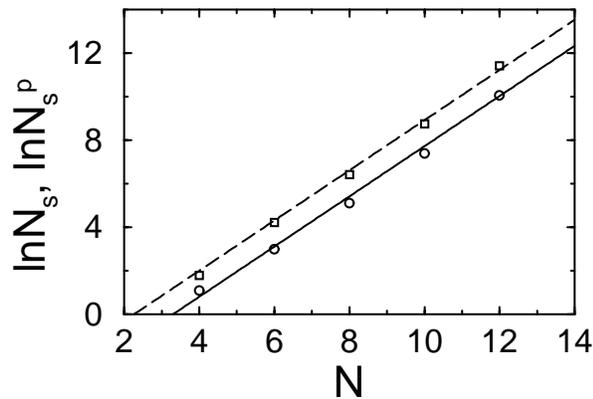}}%
\vspace{-0.15in}
\caption{\label{fig:Ns} The number of distinct mechanically stable 
packings at integer shear strain when we treat all polarizations 
the same $N_s$ (circles) and when we account for different polarizations
$N_s^p$ (squares).  The solid and long-dashed lines have slope  
$\approx 1.2$.}
\vspace{-0.12in}
\end{figure}

\section{Enumeration of MS packings at arbitrary shear strain}
\label{enumeration}

\subsection{Packing-generation protocol}
\label{Packing-generation protocol}

Zero-pressure MS packings at a given shear strain are generated using
the compression/decompression packing-generation protocol used in our
previous studies of unsheared MS packings \cite{gao}.  We briefly
outline the procedure below for completeness.  We begin the
packing-generation process by choosing random initial particle
positions within the simulation cell at packing fraction $\phi_0 =
0.50$ (which is well below the minimum packing fraction at which
frictionless MS packings occur in 2D).  We then successively increase
or decrease the diameters of the grains, with each compression or
decompression step followed by conjugate gradient minimization
\cite{numrec} of the total energy in~\eqref{total energy}.

As illustrated in Fig.\ \ref{fig:energy}, the system is decompressed
when the total energy \eqref{total energy} at a local minimum is
nonzero---{\it i.e.}, there are finite particle overlaps [cf.,
Figs.~\ref{fig:energy} \subfig{a} and \subfig{d}].  If the potential
energy of the system is zero [Fig.~\ref{fig:energy}\subfig{c}] and
gaps exist between particles [Fig.~\ref{fig:energy}\subfig{f}], the
system is compressed.  The increment by which the packing fraction
$\phi$ is changed at each compression or decompression step is
gradually decreased. Numerical details of the algorithm are provided 
in Appendix~\ref{Numerical details}.

In the final state of the packing-generation process, the potential
energy vanishes [Fig.~\ref{fig:energy}\subfig{b}], but any change of
the relative particle positions (excluding rattler particles, which
can be moved without causing particle overlap) results in an increase
in the potential energy.  Thus, the final state is a mechanically
stable configuration (or collectively jammed state \cite{stillinger})
at jamming packing fraction $\phi_J$.  At each fixed $\gamma$, MS
packings form a discrete set in configuration space.  The
packing-generation process is repeated more than $10^6$ times for at
least $100$ uniformly spaced shear strain values in the interval $0 <
\gamma < 0.5$.  A large number of independent trials is required to
enumerate nearly all MS packings because the MS probability
distribution varies by many orders of magnitude.

\begin{figure}
  \scalebox{0.35}{\includegraphics{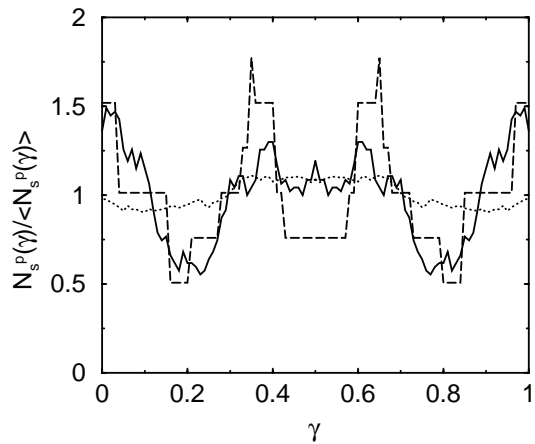}}%
  \vspace{-0.1in}
\caption{\label{fig:ns_strain} The number of distinct mechanically
stable packings $N_s^p(\gamma)$ as a function of shear strain $\gamma$
for $N=4$ (long-dashed), $6$ (solid), and $10$ (dotted).
$N_s^p(\gamma)$ is normalized by the average number of MS packings
over all shear strain $\langle N_s^p(\gamma) \rangle$.  The sampling
interval is $\Delta \gamma = 10^{-2}$.  As shown in
Fig.~\ref{fig:units_strains}, the same set of MS packings occur at
$\gamma$ and $1-\gamma$.}
\vspace{-0.12in}
\end{figure}

\begin{figure}
\scalebox{0.3}{\includegraphics{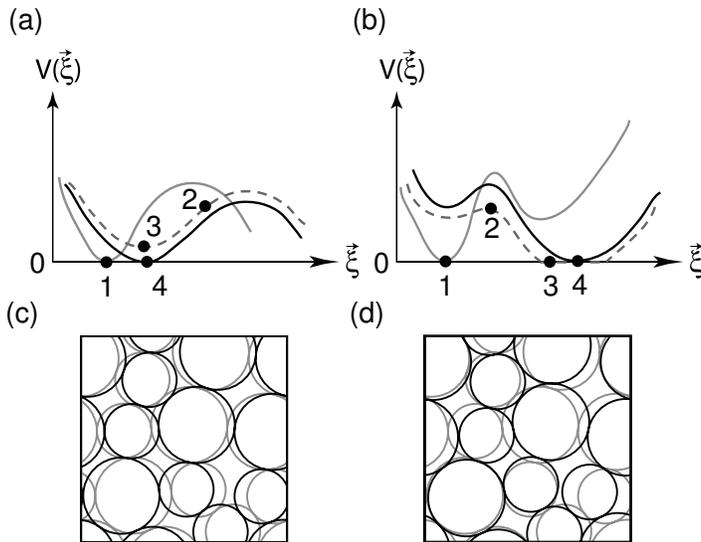}}%
\vspace{-0.0in}
\caption{\label{fig:shear} Schematic of the evolution of the
energy landscape during quasistatic shear at fixed zero pressure from
shear strain $\gamma$ to $\gamma+\delta \gamma$.  In (a), the system
evolves continuously from the local minimum at shear strain $\gamma$
(1) to the one at $\gamma + \delta \gamma$ (4) because there are no
particle rearrangement events during the shear strain interval.  In contrast,
in (b) we show that when the system undergoes a particle rearrangement 
event, it will reside in a different energy minimum at
$\gamma + \delta \gamma$ compared to the one at $\gamma$.  Snapshots of the
static packings at shear strain $\gamma$ (gray) and $\gamma + \delta
\gamma$ (black) are superimposed in (c) and (d), which correspond to the
potential energy landscape dynamics in (a) and (b), respectively.  In 
(d), three of the original contacts are removed and four new contacts 
are generated as a result of the particle rearrangements.  }
\vspace{-0.22in}
\end{figure}

\subsection{Classification of granular packings}
\label{classification}

\subsubsection{Dynamical matrix}

With our precise measurement of the jamming packing fraction $\phi_J$
to within $10^{-8}$ of the jamming point, it is very rare that two
distinct MS packings have the same $\phi_J$.  Thus, it is often
convenient to characterize MS packings by $\phi_J$.  However, in our
detailed investigations of the quasistatic evolution of sheared
granular packings, a more precise classification of MS packings is
necessary.  

To determine the set of distinct MS packings, we use the eigenvalues and
eigenvectors of the dynamical matrix \cite{barrat}
\begin{equation}
\label{dyn_matrix}
{\bf M}_{mn}=\left. \frac{\partial V}{\partial {\xi}_{m}\partial {\xi}_{n}} 
             \right|_{ {\vec \xi}_0},
\end{equation}
where $\xi_{m}$ is the $m$th component of the configuration vector
$\vec\xi$ and ${\vec \xi}_0$ gives the configuration of the reference MS
packing.  Since rattler particles do not contribute to the stability
of the system, we consider only the dynamical matrix for the mechanical
backbone of the packing.  This matrix has $d{\cal N}$ rows and
columns, where $d=2$ is the spatial dimension, ${\cal N} = N - N_r$ is
the number of particles in the mechanical backbone, and $N_r$ is the
number of rattler particles.

Since the dynamical matrix is symmetric, it has $d{\cal N}$ real
eigenvalues $\{ m_i \}$, $d$ of which are zero due to translational
invariance of the system.  In a mechanically stable disk packing, no
collective particle displacements are possible without creating an
overlapping configuration; therefore the dynamical matrix for MS
packings has exactly $d({\cal N}-1)$ positive
eigenvalues~\cite{footnote_1}.  We limit the results below to
mechanically stable packings.

\begin{figure}
\scalebox{0.5}{\includegraphics{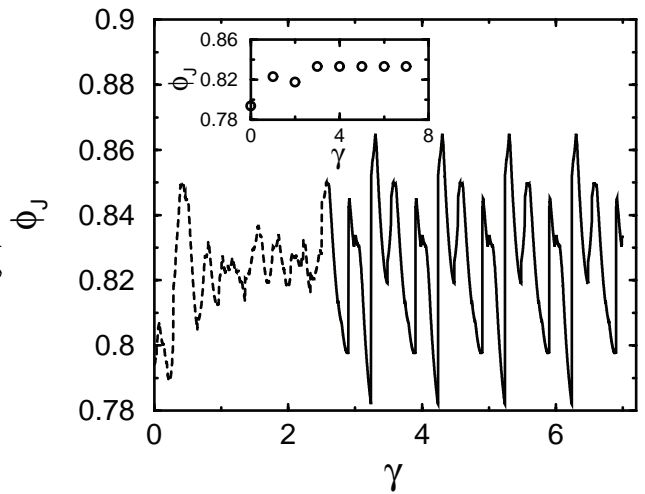}}%
\vspace{-0.1in}
\caption{\label{fig:N10_phi_gamma_1456} The jammed packing fraction
$\phi_J$ versus shear strain $\gamma$ during quasistatic shear at zero
pressure for $N=10$.  The dashed line shows the initial transient and
the solid line highlights the periodic behavior with unit period that
begins near $\gamma_t \approx 2.5$. The inset shows $\phi_J$ for the
same system in the main figure except only at integer shear strains.}
\vspace{-0.12in}
\end{figure}

\subsubsection{Polarizations}

According to our previous investigations \cite{gao,xu2}, distinct MS
packings always have distinct sets of eigenvalues $\{ m_i \}$, except
when packings can be related to each other by reflection or rotation
\cite{footnote_1}.  When we treat different `polarizations'
associated with each symmetry transformation as the same state,
distinct MS packings can be classified according to the lists of
eigenvalues of their dynamical matrices.  This approach was adopted in
Refs.~\cite{gao,xu2}, where we considered only static, isotropic
particle configurations at $\gamma=0$.  The eight equivalent
polarizations for a MS packing at $\gamma=0$ are shown in
Fig.~\ref{fig:polarization_noshear}.

In contrast, in our present study we consider continuous shear
deformations of the system.  After an isotropic unit cell is
deformed, different polarizations of a given state can be transformed
into distinct MS packings ({\it i.e.}, packings distinguishable by the lists
of eigenvalues $\{ m_i \}$).  For example, all polarizations shown in
the left (right) column of Fig.\ \ref{fig:polarization_noshear}, after
a step strain are transformed into non-equivalent configurations.  Our
present classification scheme for MS packings takes this effect into
account.

Accordingly, we treat states that differ by reflection or rotation by
the angle $\alpha=\pi/2$ or $3\pi/2$ as distinct MS packings.
However, due to symmetry of shear flow (cf.\ bottom of Fig.\
\ref{fig:polarization_noshear}), the states that are related by a
rotation by the angle $\alpha=\pi$ deform in equivalent way.
Therefore, such states are treated as equivalent states. Further
details regarding polarizations of MS packings, including the
procedure we use to distinguish polarizations, are discussed in
Appendix~\ref{Appendix: polarizations}.

\subsection{Distinct MS packings at integer and non-integer strains}

MS packings at integer shear strains have equivalent boundary
conditions to those in standard square periodic cells (as shown in
Fig.~\ref{fig:units_strains}).  Thus, we can use our previous
extensive calculations of MS packings generated in square cells with
periodic boundary conditions \cite{gao}.   

In Table~\ref{tab:table5} we show the number of distinct MS packings
at integer shear strains $N_s$ when we treat all polarizations as the
same (data from Ref.~\cite{gao}) and $N_s^p$ when we account for
different polarizations (as described above).  The ratio $N_s^p/N_s$
is smaller than $4$ because of reflection or rotation symmetry of some
MS packings.  As depicted in Fig.\ \ref{fig:Ns}, both $N_s$ and
$N_s^p$ grow exponentially with system size.

The number of distinct MS packings $N_s^p(\gamma)$ (including
polarizations) versus shear strain is shown in
Fig.~\ref{fig:ns_strain} for several system sizes.  The results show
that (1) the maximum number of distinct packings occurs at integer and
half-integer shear strains, and (2) there is a noticeable dip in
$N_s^p$ at $\gamma^* \approx 0.2$.  However, as the system size
increases, the number of distinct MS packings becomes uniform as a
function of shear strain.

\begin{table}
\caption{\label{tab:orbit_data} The maximum transient shear strain
$\gamma_t^{\rm max}$ and period $T$ of each distinct periodic orbit in
integer strain units for several system sizes $N$.  ($\gamma_t^{\rm
max}$ is rounded to the nearest integer shear strain.)  In the fourth
column, we list the average jammed packing fraction $\langle \phi_J \rangle$ 
of the MS packings that are
dynamically accessible at large shear strain for each each periodic
orbit.  For systems with multiple periodic orbits, we list
$\gamma^{\rm max}_t$, $T$, and $\langle \phi_J \rangle$ for each.}
\begin{ruledtabular}
\begin{tabular}{rrrr}
$N$ & $\gamma^{\rm max}_t$ & $T$ & $\langle \phi_J \rangle$\\
\tableline
$4$ & $2$ & $1$ & $0.829$\\ 
\tableline
$6$ & $2$ & $1$ & $0.777$\\
\tableline
$8$ & $5$ & $1$ & $0.820$\\
 & $1$ & $2$ & $0.825$\\
\tableline
$10$ & $1$ & $1$ & $0.815$ \\
 & $3$ & $1$ & $0.808$ \\
 & $4$ & $1$ & $0.833$ \\ 
 & $2$ & $2$ & $0.812$ \\
\tableline
$12$ & $1$ & $1$ & $0.782$ \\
 & $12$ & $3$ & $0.815$ \\
\tableline
$14$ & $0$ & $1$ & $0.811$ \\
 & $1$ & $1$ & $0.789$ \\
 & $2$ & $1$ & $0.814$ \\
 & $2$ & $1$ & $0.816$ \\
 & $2$ & $1$ & $0.814$ \\
 & $3$ & $1$ & $0.847$ \\
 & $3$ & $1$ & $0.815$ \\
 & $4$ & $1$ & $0.847$ \\
 & $36$ & $1$ & $0.836$ \\
 & $8$  & $5$ & $0.804$ \\
\end{tabular}
\end{ruledtabular}
\end{table}

\begin{table}
\caption{\label{tab:table4} Catalog of the $N_s^p=68$ possible
transitions from a MS packing with index $i$ at 
$\gamma=0$ to index $j$ at $\gamma=1$ for quasistatic shear flow at zero
pressure for $N=6$.}
\begin{ruledtabular}
\begin{tabular}{rrrrcc}
$i(\gamma=0)$ & $j(\gamma=1)$ & $i(\gamma=0)$ & $j(\gamma=1)$ & $i(\gamma=0)$ 
& $j(\gamma=1)$\\
\tableline
$1$ & $37$ & $26$ & $67$ & $51$ & $62$\\ 
$2$ & $67$ & $27$ & $37$ & $52$ & $67$\\
$3$ & $37$ & $28$ & $62$ & $53$ & $67$\\
$4$ & $62$ & $29$ & $67$ & $54$ & $67$\\
$5$ & $67$ & $30$ & $62$ & $55$ & $67$\\
$6$ & $62$ & $31$ & $62$ & $56$ & $67$\\ 
$7$ & $67$ & $32$ & $67$ & $57$ & $67$\\
$8$ & $67$ & $33$ & $62$ & $58$ & $67$\\
$9$ & $62$ & $34$ & $37$ & $59$ & $67$\\
$10$ & $37$ & $35$ & $67$ & $60$ & $67$\\
$11$ & $67$ & $36$ & $62$ & $61$ & $67$\\ 
$12$ & $67$ & $37$ & $67$ & $62$ & $67$\\
$13$ & $67$ & $38$ & $62$ & $63$ & $67$\\
$14$ & $67$ & $39$ & $37$ & $64$ & $62$\\
$15$ & $67$ & $40$ & $67$ & $65$ & $67$\\
$16$ & $62$ & $41$ & $37$ & $66$ & $67$\\ 
$17$ & $62$ & $42$ & $67$ & $67$ & $67$\\
$18$ & $67$ & $43$ & $67$ & $68$ & $67$\\
$19$ & $62$ & $44$ & $62$ & $$ & $$\\
$20$ & $37$ & $45$ & $67$ & $$ & $$\\
$21$ & $62$ & $46$ & $37$ & $$ & $$\\ 
$22$ & $67$ & $47$ & $67$ & $$ & $$\\
$23$ & $67$ & $48$ & $67$ & $$ & $$\\
$24$ & $67$ & $49$ & $67$ & $$ & $$\\
$25$ & $37$ & $50$ & $37$ & $$ & $$\\
\end{tabular}
\end{ruledtabular}
\end{table}

\section{Quasistatic shear flow at zero pressure}
\label{Quasistatic evolution}

\subsection{System dynamics}

\subsubsection{Quasistatic shear-strain steps}
\label{Quasistatic shear-strain steps}

To mimic quasistatic evolution of a frictionless granular packing, we
apply a sequence of successive shear-strain steps of size $\delta
\gamma \ll 1$ to a system of bidisperse disks with shear-periodic 
boundary conditions.  Each step of the protocol consists of
(1) shifting the $x$-coordinate of the particles,
\begin{equation}
\label{strain}
x_i \rightarrow x_i + \delta \gamma y_i,
\end{equation}
in conjunction with the corresponding deformation of the unit cell;
and (2) the compression/decompression packing-generation process (described in
Sec.\ \ref{Packing-generation protocol}) to achieve a zero-pressure MS
configuration with infinitesimal particle overlaps.

This procedure generates quasistatic shear flow at constant {\it zero
pressure} (but varying packing fraction), which is an appropriate description
of slowly sheared granular matter, where particle overlaps are always
minimal.  During the evolution, the system constantly expands and
contracts to remain at the onset of jamming with packing fraction
$\phi_J$ that depends on $\gamma$.  We note that our procedure is
distinct from the quasistatic evolution used in recent investigations
of sheared glass-forming liquids \cite{osborne,isner}, where a
constant-volume ensemble was implemented.

\subsubsection{Particle rearrangements}

During a single shear-strain step $\delta\gamma$, particle positions
can either change continuously or a sudden particle rearrangement can
occur.  A continuous evolution step is schematically depicted in the
left panels of Fig.\ \ref{fig:shear}, and the right panels illustrate
a strain step that yields a particle rearrangement.

In both cases, the system initially resides in the local minimum (point
$1$) in the energy landscape.  Since this minimum corresponds to a MS
packing, its energy is infinitesimal.  When the affine
transformation~\eqref{strain} is applied, the system moves to point
$2$ in the energy landscape.  In the case of continuous evolution,
depicted in Fig.\ \ref{fig:shear}\subfig{a}, the energy minimization
(point $3$) and subsequent decompression (point $4$) move the system
back to the neighborhood of the initial packing.  As illustrated in
\ref{fig:shear} \subfig{c}, at the initial and final points ($1$) and
($4$), the topology of contact networks is unchanged.

In contrast, when the energy minimization that follows an affine
shear-strain step drives the system into a zero-energy region
corresponding to an unjammed packing ({\it i.e.} point $3$ in Fig.\
\ref{fig:shear}\subfig{b}), upon subsequent compaction the system is
driven to a MS packing (point $4$) with a different contact network
than the initial packing (point $1$), as illustrated in Fig.\
\ref{fig:shear}\subfig{d}.

\begin{figure}
\scalebox{0.35}{\includegraphics{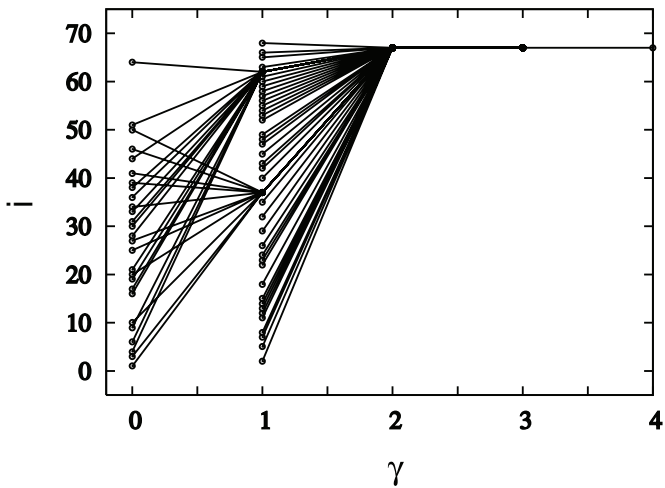}}%
\vspace{-0.1in}
\caption{\label{fig:tree} This `tree' diagram for $N=6$ shows the
evolution at integer shear strains of the system undergoing
quasistatic shear flow at zero pressure.  The system is initialized at
all $68$ possible MS packings using the same indexes $i$ as in
Table~\ref{tab:table4}. The shear strain axis has been shifted so that
the periodic orbits for all initial MS packings begin at $\gamma=2$.
There is only one MS packing ($67$) that is dynamically accessible at
large shear strain.}
\vspace{-0.12in}
\end{figure}

During the continuous portion of the quasistatic shear strain
evolution, the MS packings that are visited do not depend on the
energy minimization method (e.g., energy relaxation via dissipative
forces versus the conjugate gradient algorithm~\cite{gao}) or
parameters related to the compaction and decompression processes.
Dynamical features do not influence the MS packings that are obtained
along the continuous region because the system remains in the basin of
the original mechanically stable packing.  However, when a shear
strain step leads to particle rearrangements, the system is taken to a
new region of the energy landscape, and the energy minimization
scheme, compaction and decompression rates, and even the location of
rattler particles can influence the final MS packing.  In small
systems ($N<16$), we find that the dynamics is weakly sensitive to
these features, whereas in larger systems noise and protocol
dependence play an important role in determining steady-state MS
packing probabilities. In Sec.~\ref{random splitting}, we will
describe future work in which we will tune the dynamics to determine its 
influence on the transition rates among MS packings.

\subsection{Deterministic and contracting evolution of small frictionless 
MS packings}
\label{contraction}

Our key results regarding quasistatic evolution of small granular
systems are summarized in Figs.\ \ref{fig:N10_phi_gamma_1456} and
\ref{fig:tree} and in Tables \ref{tab:orbit_data} and \ref{tab:table4}.
The results show that (1) the evolution of small systems is
deterministic; (2) after transient evolution, the system becomes
locked into a periodic orbit; and (3) the evolution in
configuration space is strongly contracting, in the sense that each
unit strain leads to a significant reduction of the number of
dynamically accessible MS packings.

\paragraph*{Deterministic evolution}

The deterministic character of the system evolution over continuous
portions of the trajectories can be justified using arguments based on
the continuity of the energy landscape, similar to those
illustrated in Fig.\ \ref{fig:shear} \subfig{a}.  The exceptions are
bifurcation points, which will be described in Sec.\ \ref{random splitting}.
Indeterminacy can also occur due to the presence of rattlers or random
particle motions in unjammed configurations.  For systems with
$N < 16$ evolved according to the algorithm described in Sec.\
\ref{Quasistatic shear-strain steps}, the observed evolution was always
completely deterministic. Random evolution in larger systems and
systems with random noise are discussed in Sec.\ \ref{random
splitting}.

\paragraph*{Periodic orbits}

In Fig.~\ref{fig:N10_phi_gamma_1456}, we track the evolution of the
jammed packing fraction $\phi_J$ as a function of shear strain
$\gamma$ during zero-pressure quasistatic shear flow after the system
is initialized in one of the MS packings at $\gamma=0$ for $N=10$
particles.  Fig.~\ref{fig:N10_phi_gamma_1456} shows the complete
trajectory and the inset shows $\phi_J$ only at integer strains.  We
observe that after a short transient of approximately two units of
strain, the system becomes locked into a periodic orbit (with unit
period $T=1$).  Similar behavior is observed for systems with $N=4$ to
$14$ particles, as summarized in Table \ref{tab:orbit_data}.  Although
we have a limited range of system sizes for which a complete analysis
of the quasistatic evolution has been preformed, the results show that
both the transient time $\gamma_t$ and period $T$ increase somewhat
with system size. (Note that several periodic orbits for $N=12$ and
$14$ particles have anomalously large transients, cf. Sec.~\ref{random
splitting}.)

\paragraph*{Contracting evolution}

The contracting character of the quasistatic shear flow is illustrated
in the `tree' diagram in Fig.\ \ref{fig:tree}.  This digram represents
the complete set of trajectories, shown for integer strain, for a
system of $N=6$ particles.  For this system size there are $N_s^p=68$
possible MS packings at any given integer shear strain, yet only {\it
one} of them is dynamically accessible in the large shear strain
limit.

The contraction of the set of dynamically
accessible states as the system evolves results from the deterministic
and irreversible character of quasistatic evolution.  As illustrated
in Fig.\ \ref{fig:tree}, from each MS packing at an integer strain
$\gamma$, the system transitions to a unique MS packing at the strain
$\gamma+1$; however, more than one state can transition to a given
state.  A complete catalog of transitions at integer shear strains for
the system of $N=6$ particles is given in Table \ref{tab:table4}.
From this table, a complete list of trajectories of the whole
evolution process can be constructed {\it without further
simulations}.

As shown in Table \ref{tab:orbit_data}, the number of periodic orbits
to which the system contracts increases with the system size.
However, for all systems studied, the number of orbits and the number
of states visited at long times is small compared to the total number of
MS packings.

The results in the last column of Table \ref{tab:orbit_data} give the
average jammed packing fraction $\langle \phi_J\rangle$ of the MS
packings sampled for a given periodic orbit. The MS packings that
occur in periodic orbits typically have the highest packing fractions
out of the entire distribution.

\section{Family structure of the set of MS packings}
\label{families}

An analysis of the deterministic evolution of the system monitored at integer
values of strain is sufficient to conclude that the dynamics is
contracting and periodic at large shear strains.  However, it might be puzzling
that both the transient shear strain and period are so short, {\it i.e.}, much
shorter than the number of MS packings (cf., Tables \ref{tab:table5}
and \ref{tab:orbit_data}).  To shed light on this behavior we will now
analyze how the sets of MS packings at different values of shear strain are
connected.

\begin{figure}
\scalebox{0.23}{\includegraphics{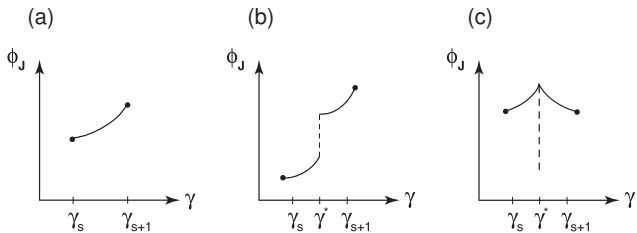}}%
\vspace{-0.0in}
\caption{\label{fig:evolution} Schematic of the evolution of the
jamming packing fraction $\phi_J$ during the shear strain interval
$\gamma_s$ to $\gamma_{s+1}$.  In (a) the particle contact network
does not change from $\gamma_s$ to $\gamma_{s+1}$, while
there is a discontinuity in $\phi_J$ in (b) and in the derivative 
of $\phi_J$ in (c) at $\gamma^*$, which signal a change in the contact 
network.}
\vspace{-0.22in}
\end{figure}

\subsection{Construction of continuous geometrical families of MS packings}
\label{construction}

As described in Sec.~\ref{classification}, we can identify the distinct 
MS packings at a given shear strain using the eigenvalues of the 
dynamical matrix.  However, this method will not work for comparing 
MS packings at different shear strains since the eigenvalues vary continuously 
with $\gamma$.    

To study the relationship between distinct MS packings at different
shear strains, we divide the shear strain region into small intervals
$\gamma_{s+1} - \gamma_{s}$.  For each distinct MS packing at
$\gamma_s$, we monitor the particle contact network as the system
evolves toward $\gamma_{s+1}$ (and $\gamma_{s-1}$).  In
Fig.~\ref{fig:evolution} (a), we show the continuous evolution of
$\phi_J$ between $\gamma_s$ and $\gamma_{s+1}$, which implies that
there are no rearrangement events and no changes in the particle
contact network during this interval.  Thus, the continuous evolution
of $\phi_J$ identifies a portion of a {\it geometrical family} of MS
packings all with the same particle contact network that exist over a
continuous shear strain interval.

In our simulations, we find that changes in the network of particle
contacts ({\it i.e. switches from one geometrical family to another})
is accomplished either by jumps (discontinuities in $\phi_J$) or
kinks (discontinuities in the derivative of $\phi_J$ with respect to
$\gamma$) as shown in Figs.~\ref{fig:evolution} \subfig{b} and
\subfig{c}, respectively.  As will be discussed in Sec.~\ref{map}, a
discontinuity in $\phi_J$ corresponds to a system instability and a
change in the contact network with finite particle
displacements; whereas a discontinuity in the derivative of $\phi_J$
corresponds to a change of the contact network without finite particle
displacements.
 
To construct the complete map of distinct geometrical families for all
shear strain, we can simply link the equivalent geometrical families
at the shear strain endpoints $\gamma_s$, or terminate the family if
it has no counterpart.  Because of the contracting dynamics described
in Sec.~\ref{contraction}, it is important to use sufficiently small
shear strain intervals so that we have enough shear strain
resolution to capture small families. (See Appendix~\ref{Numerical
details} for additional details for constructing geometrical
families.)

\begin{figure}
  \scalebox{0.45}{\includegraphics{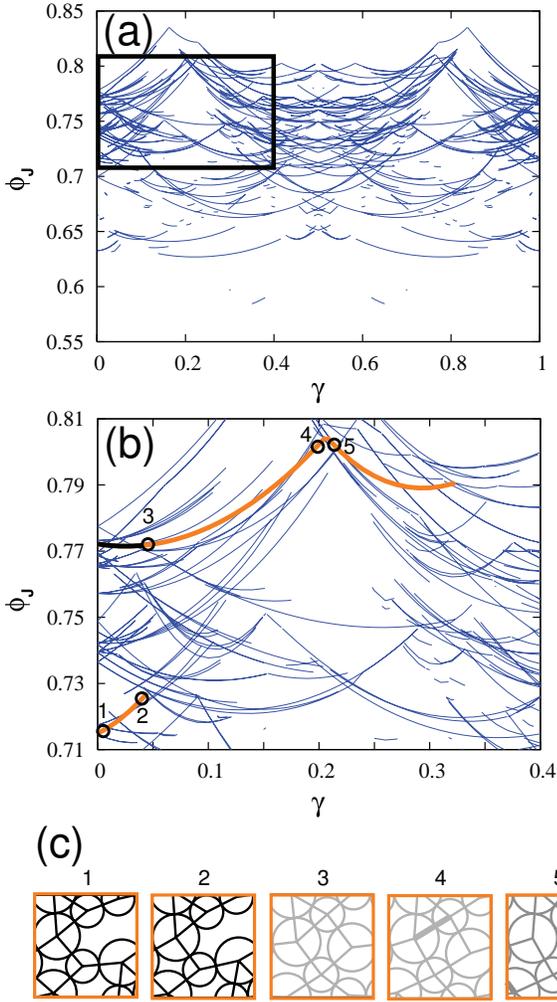}}%
  \vspace{-0.05in}
  \caption{\label{fig:family_evolution} (Color online) \subfig{a}
Complete geometrical family map---jamming packing fraction
$\phi_J(\gamma)$ as a function of shear strain $\gamma$---for $N=6$
(blue lines). \subfig{b} Close-up of boxed region in \subfig{a}.  The
orange (black) lines indicate increasing (decreasing) shear strain
evolution. \subfig{c} Particle configurations at five points during
evolution within the family map are also shown.  Solid lines
connecting particle centers represent particle contacts; each distinct
network is given a different grayscale. Contacts denoted by thick
lines in panels $4$ and $5$ are either gained or lost as the system
evolves from configuration $4$ to $5$.}
\vspace{-0.22in}
\end{figure}

\begin{table}
\caption{\label{tab:table6} Statistics for complete geometrical family
maps. Total number of distinct geometrical families $N_f$, average
number of distinct MS packings $\langle N_s^p(\gamma)\rangle$, average
packing fraction $\langle \phi_J(\gamma) \rangle$, average family
length $\xi$, and average family second derivative ${\cal C}$ for
several system sizes $N$~\cite{foot}.}
\begin{ruledtabular}
\begin{tabular}{rrrrcc}
$N$ & $N_f$ & $\langle N_s^p(\gamma)\rangle$ & $\langle \phi_J(\gamma) 
\rangle$ & $\xi$ & ${\cal C}$\\
\tableline
$4$ & $15$ & $4$ & $0.788$ & $0.18$ & 1.7\\ 
$6$ & $334$ & $47$ & $0.739$ & $0.06$ & 5.6\\
$10$ & $34822$ & $2896$ & $0.753$ & $0.03$ & 4.9\\
\end{tabular}
\end{ruledtabular}
\end{table}

\begin{figure}
\scalebox{0.35}{\includegraphics{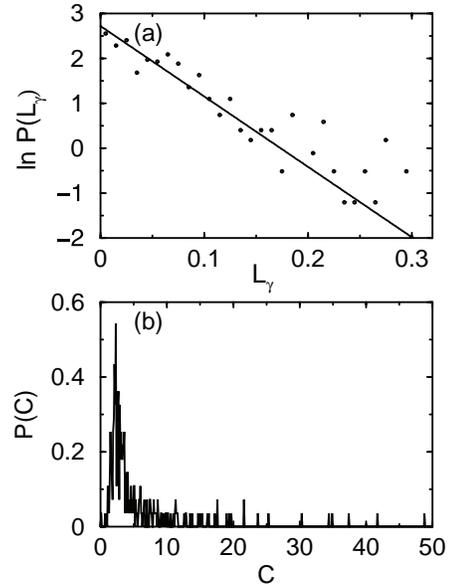}}%
\vspace{-0.18in}
\caption{\label{fig:binned_family_length} (a) Probability distribution
$P(L_\gamma)$ for the length (in units of strain) of geometrical
families for $N = 6$.  The solid line represents exponential decay
with decay length $\xi \sim 0.06$. (b) Distribution $P({\cal
C})$ of second derivatives of $\phi_J(\gamma)$ (with respect to shear
strain $\gamma$) for the distinct geometrical families for $N=6$. The
family average is ${\cal C} \sim 5.6$.}
\vspace{-0.22in}
\end{figure}

\subsection{Complete map of MS packings for continuous
shear strain}
\label{map}

We find that a particularly simple, pictorial method to distinguish
geometrical families is by monitoring the jamming packing fraction
$\phi_J$ as a function of shear strain (instead of a complete
representation of the particle contact network).  The complete map of
MS packings---$\phi_J$ for all distinct MS packings at all shear
strains---is displayed in Fig.~\ref{fig:family_evolution} for $N=6$.
The structure of the map is quite complex even for $N=6$, and it
possesses a number of distinctive features.

First, the map is composed of a finite number ($N_f=334$ for $N=6$) of
curved concave-up segments each of which corresponds to a distinct
geometrical family of MS packings.  Second, the parabola-like curves
either end discontinuously [cf., point 2 in the blowup shown in Fig.\
\ref{fig:family_evolution} \subfig{b}] or form a kink [points 4 and 5
in Fig.\ \ref{fig:family_evolution} \subfig{b}].  Third, the curves
significantly vary in length, and they are not symmetric about the
apexes (which are distributed over the full range of $\gamma$
\cite{footnote3}).  

We find that a given parabola-like curve ends when either the contact
network of that particular continuous region becomes unstable and the
system jumps to a new one or the contact network merges with another
network to form a kink.  Examples of contact networks corresponding to
characteristic points on the family map are shown in 
Fig.\ \ref{fig:family_evolution} \subfig{c}.

Figure \ref{fig:binned_family_length} shows that the family-length
distribution is exponential (with the characteristic strain of
$\xi \sim 0.06$ for N=6 \cite{footnote2}) and the distribution of second
derivatives possesses a strong peak.  We speculate that the decay
length $\xi$ is related to the average yield strain in frictionless MS
packings.  Thus, it is important to study $\xi$ as a function of
system size and this direction will be pursued in future work.

The jammed packing fraction $\phi_J$ has parabolic-like dependence on
$\gamma$ because we consider jammed disk packings. The general feature
of these packings is that as shear strain first increases the system
must dilate (packing fraction decreases) as particles climb over each
other~\cite{reynolds}.  However, beyond a critical shear strain the
system must compact to maintain particle contacts.  

Since the shape of the family map $\phi_J=\phi_J(\gamma)$ is entirely
determined by the geometry of the particle contacts at zero pressure,
this shape is independent the detailed form of the finite-range
potential~\eqref{spring potential}.  In particular, it does not depend
on the power $\alpha$ defining the interparticle elastic forces.

We note that recent simulations have found that packings of
ellipsoidal particles possess a large number of low-energy vibrational
modes, which are quartically, not quadratically,
stabilized~\cite{donev,mailman,zeravcic}. Thus, we predict
non-parabolic dependence of $\phi_J(\gamma)$ for small quasistatically
sheared packings of ellipsoidal particles.

In Table~\ref{tab:table6} we summarize our results by compiling
statistics of the complete geometrical family maps for several system
sizes $N$.  We provide the total number of distinct geometrical
families $N_f$, average number of distinct MS packings $\langle
N_s^p(\gamma)\rangle$, average packing fraction $\langle
\phi_J(\gamma) \rangle$, average family length $\xi$, and average
family second derivative ${\cal C}$ (with respect to $\gamma$), where
$\langle .\rangle$ indicates an average over shear strain $\gamma$.

\begin{figure}
  \scalebox{0.4}{\includegraphics{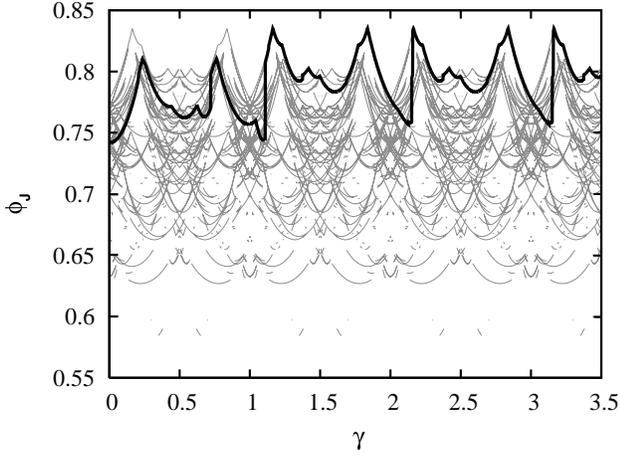}}%
\vspace{-0.1in}
\caption{\label{fig:single_state} The complete
geometrical family map as shown in Fig.~\ref{fig:family_evolution} as
a function of shear strain $\gamma$ for $N=6$ (gray lines).  The black
line shows the evolution of $\phi_J$ during quasistatic shear strain
starting from a single MS packing with $\phi_J=0.742$ at $\gamma=0$.  }
\vspace{-0.15in}
\end{figure}

\begin{figure}
  \scalebox{0.4}{\includegraphics{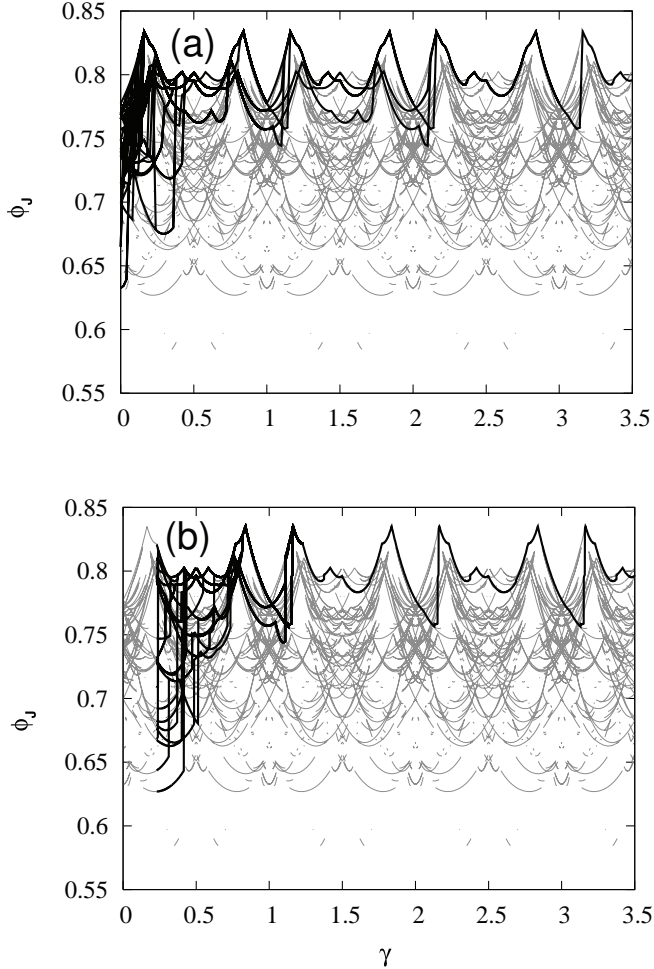}}%
\vspace{-0.1in}
\caption{\label{fig:families_definition} The complete
geometrical family map as shown in Fig.~\ref{fig:family_evolution} as
a function of shear strain $\gamma$ for $N=6$ (gray lines).  In (a)
and (b), we show the evolution of $\phi_J(\gamma)$ under quasistatic 
shear strain starting from all distinct MS packings at
$\gamma=0$ and $0.23$, respectively (black lines).  The number of distinct 
MS packings at $\gamma=0.23$ is smaller than the number at $0$ by a 
factor of $\approx 3$.}
\vspace{-0.15in}
\end{figure}

\begin{figure}
  \scalebox{0.4}{\includegraphics{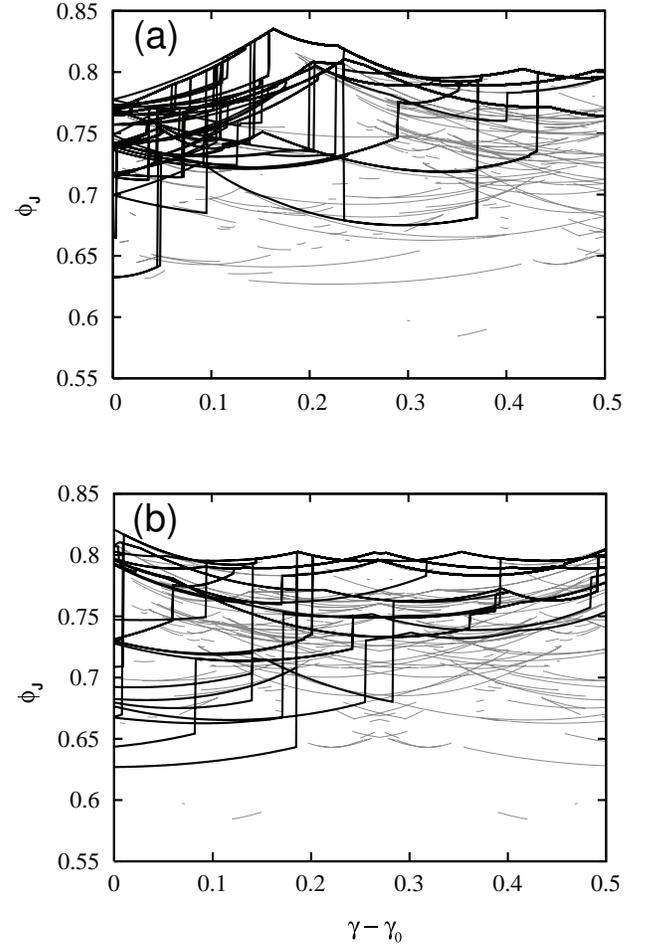}}%
\vspace{-0.1in}
\caption{\label{fig:closeup_families_definition} Magnifications of the
complete family maps and quasistatic shear evolution at
$\gamma-\gamma_0$ for initial shear strains \subfig{a} $\gamma_0=0$
and \subfig{b} $0.23$ from Figs.~\ref{fig:families_definition}
\subfig{a} and \subfig{b}.  }
\vspace{-0.15in}
\end{figure}

\begin{figure}
\scalebox{0.36}{\includegraphics{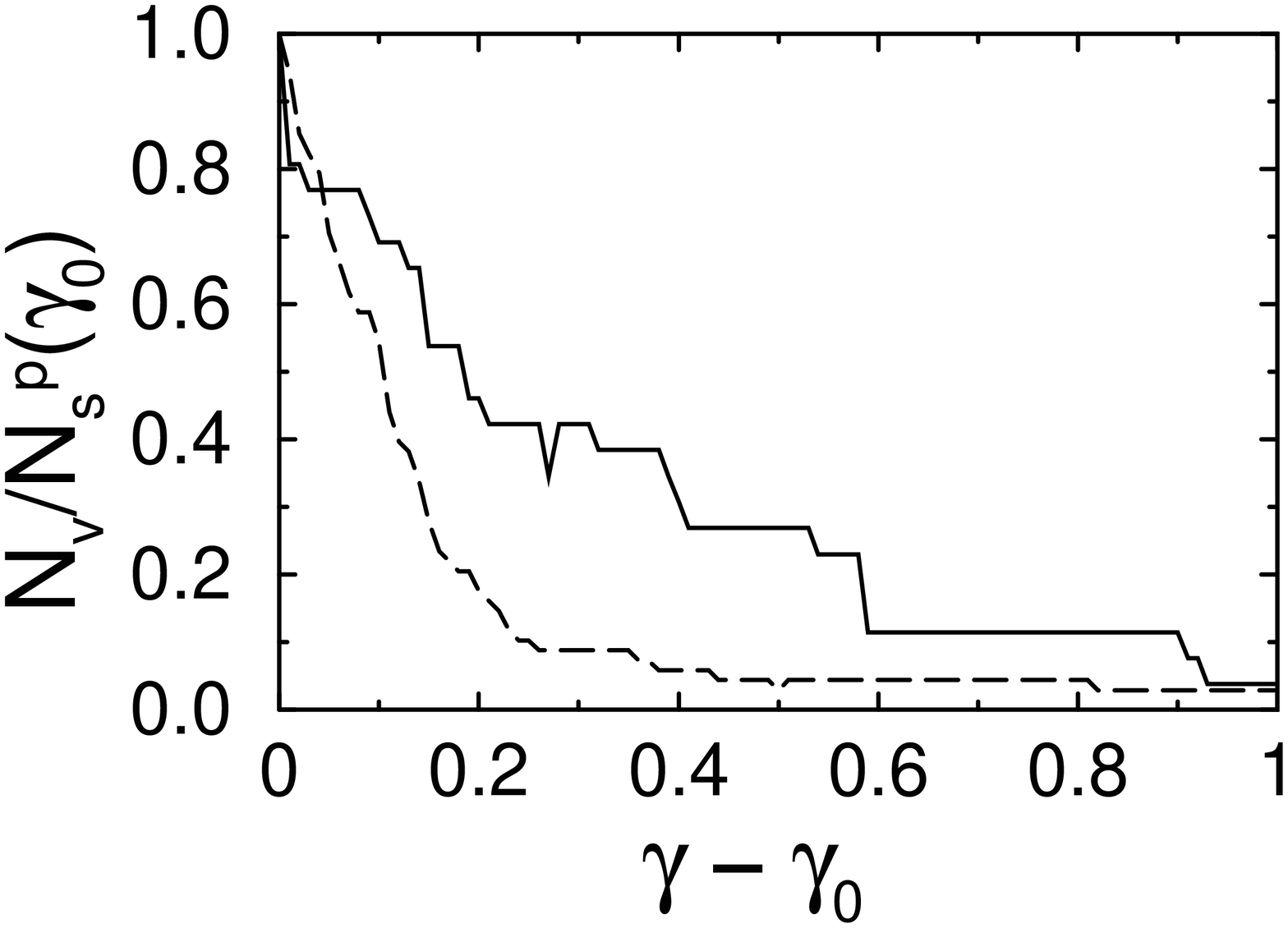}}%
\vspace{-0.18in}
\caption{\label{fig:jump} The number $N_v$ of distinct MS packings
visited during quasistatic shear (normalized by $N_s^p(\gamma_0)$) versus
$\gamma-\gamma_0$ for $N=6$ obtained obtained by summing over all of
the possible MS packings initialized at $\gamma_0$ for $\gamma_0=0$
(dashed line) and $0.23$ (solid line).}
\vspace{-0.22in}
\end{figure}

\subsection{Quasistatic evolution}
\label{evolve}

Since we have now mapped all of the MS packings over the full range of
shear strain, we can now relate key features of the quasistatic
evolution at zero pressure to the family structure of the MS packings.
We recall from Sec.\ \ref{Quasistatic evolution} that the evolution is
strongly contracting, non-ergodic, and after a short transient, the
system settles on a periodic trajectory.

\subsubsection{Topological changes and transitions between geometrical 
families of MS packings}
\label{Topological changes and transitions between families of 
MS packings}

We first examine a detailed snapshot of the evolution depicted by
orange line in Fig.\ \ref{fig:family_evolution} \subfig{b}.
Characteristic points along the trajectory are indicated by the
circles marked 1--5, and the corresponding contact networks of the
evolving MS packing are shown in \subfig{c}.

The evolution starts at point 1, which represents one of the MS packings in
a system with an unstrained unit cell.  When the strain is gradually
increased, the MS packing evolves continuously from point $1$ to $2$
along the geometrical family that includes the initial point.  As shown in
\subfig{c}, there is no topological change of the contact
network during the continuous part of the trajectory.

When the strain is increased beyond point 2 (where the family ends),
the system discontinuously transitions to another branch of MS
packings.  During the transition, the packing fraction $\phi_J$
discontinuously increases, particles undergo finite displacements, and
a new contact network is formed.  Due to the condition that the system
must be mechanically stable at zero pressure, only jumps up in the
packing fraction are allowed.

With further increases in shear strain, the evolution remains
continuous until point $4$, where the system encounters a kink in
$\phi_{J}(\gamma)$ ({\it i.e.}, discontinuity in the derivative of
$\phi_J(\gamma)$).  The kink signals a change in the particle contact
network, with no finite particle displacements.  Such a change occurs
when a new interparticle contact is formed and another is broken to
prevent the system from being overconstrained, as shown in panels $4$
and $5$ in Fig.~\ref{fig:family_evolution} \subfig{c}.  A change in
the direction of particle motion also accompanies a kink in $\phi_J$.
Note that kinks provide an important mechanism by which the $\phi_J$
of a MS packing can decrease during quasistatic shear at zero
pressure. Continued increases of the shear strain beyond point $5$
give rise to continuous evolution of $\phi_{J}(\gamma)$ until the next
discontinuity.  Only two types of discontinuities in $\phi_J$---jumps
and kinks---have been observed.

We note that jumps of the system to new contact networks at terminal
points of continuous families provide a mechanism for hysteresis and
irreversibility as found in recent experiments of cyclically sheared
granular systems~\cite{nicolas}.  If the applied shear strain is
reversed at point $3$ in Fig.~\ref{fig:family_evolution}~\subfig{b},
the system will evolve along the black-highlighted region of
$\phi_J(\gamma)$, not the original one sampled during the increasing
applied shear strain.

For small systems ($N<16$) undergoing quasistatic shear strain at zero
pressure (using conjugate gradient energy minimization), we have found
that the process of jumping from an old to a new MS packing family is
deterministic, but further study is required to determine to what
extent it depends on the packing generation protocol.  Below, we
will show that bifurcations in configuration space can affect the
destination of MS packings during quasistatic shear strain; this topic
will be discussed in more detail in Sec.~\ref{random splitting}.

\subsubsection{Periodic orbits and contracting evolution}
\label{Periodic orbits and contracting evolution}

A representative trajectory $\phi_J=\phi_J(\gamma)$ over several
strain units is shown in Fig.~\ref{fig:single_state}.  The system is
initialized in one of the MS packings ($\phi_J=0.742$) at $\gamma=0$,
and we plot $\phi_J$ as a function $\gamma$ overlayed on the complete
family map for $N=6$.  The trajectory exhibits continuous parts
separated by kinks and jumps, similar to those described in Sec.\
\ref{Topological changes and transitions between families of MS
packings}.  After a short transient evolution for $\gamma<\gamma_t
\approx 1.25$, the trajectory becomes periodic with period $1$,
consistent with the results discussed in Sec.\ \ref{contraction}.

The behavior shown in Fig.~\ref{fig:single_state} is typical for 
small packings under quasistatic shear strain.  
In Fig.~\ref{fig:families_definition} \subfig{a}, we show the
evolution of $\phi_J$ for all $N_s^p(\gamma_0=0)=68$ MS packings at
initial shear strain $\gamma_0=0$.  After a short initial transient
$\gamma \approx 0.4$, most of the initial conditions have converged
onto one of several persistent trajectories.  After $\gamma_t \approx
2.25$, all trajectories have converged onto a periodic orbit with unit
period. 

Figure \ref{fig:families_definition} \subfig{b} shows the
corresponding results for all $N_s^p(\gamma_0=0.23)=26$ MS packings
beginning at $\gamma_0 \approx 0.23$ (which is a region of shear
strain at which there are the smallest number of MS packings according
to the results depicted in Fig.~\ref{fig:ns_strain}).  Qualitatively,
the picture is similar to that in \subfig{a}: there is rapid
contraction, several persistent trajectories, and then collapse onto a
single periodic orbit.  The two ensembles with $\gamma_0=0$ and $0.23$
sample different MS packings during the transient, but evolve to the
same periodic orbit found in \subfig{a}.  Thus, these systems sample
only a small fraction of the possible geometrical families in the large
shear strain limit.

To determine the source of the rapid contraction of the number of
dynamically accessible states, we magnify the region of small shear
strains $\gamma-\gamma_0$ in Fig.\
\ref{fig:closeup_families_definition}.  The results indicate that the
contraction occurs when more than one trajectory jumps to the same branch.
The initial contraction happens quickly as depicted in Fig.\
\ref{fig:jump}, although there is some dependence of the contraction 
on $\gamma_0$.  

An examination of the closeup in Fig.\
\ref{fig:closeup_families_definition} suggests that the source of the
quick contraction and transient to the final periodic orbit is
threefold: (i) there is a large number of short families (cf., Fig.\
\ref{fig:binned_family_length}); (ii) jumps always occur up in packing
fraction, which makes low-packing-fraction MS packings dynamically
inaccessible; and (iii) some families collect significantly more
MS packings per unit length than the others, as discussed below.

\begin{figure}
\scalebox{0.3}{\includegraphics{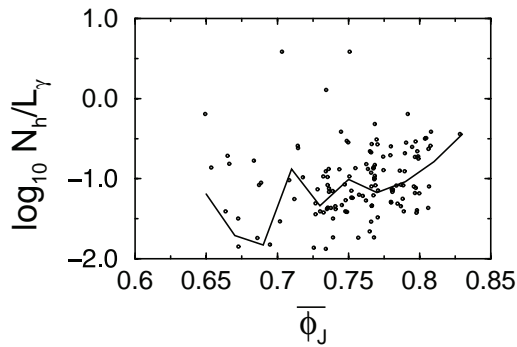}}%
\vspace{-0.18in}
\caption{\label{fig:collect} The number of trajectories $N_h$ that
jumped to a particular geometrical family (normalized by the geometrical
family length) during quasistatic shear versus the average packing
fraction ${\overline \phi}_J$ of the given geometrical family.  Jumps
were collected over narrow strain intervals $\Delta \gamma = 10^{-2}$
for $N=6$. The solid line indicates an average over bins in 
packing fraction with size $0.02$.}
\vspace{-0.22in}
\end{figure}

To estimate the `basins of attraction' for each geometrical family, we
calculated the number of jumps that each geometrical family collects
during quasistatic shear strain.  In Fig.~\ref{fig:collect}, we show
the number of trajectories $N_h$ that jump to a particular geometrical
family during quasistatic shear as a function of the average jammed
packing fraction ${\overline \phi_J}$ of the geometrical family for
$N=6$.  $N_h$ is normalized by the family length to account for the
fact that longer families can collect more trajectories.  As shown in
Fig.~\ref{fig:collect} we find a general trend that families at higher
packing fractions collect relatively more of the trajectories.

\begin{figure}
\scalebox{0.32}{\includegraphics{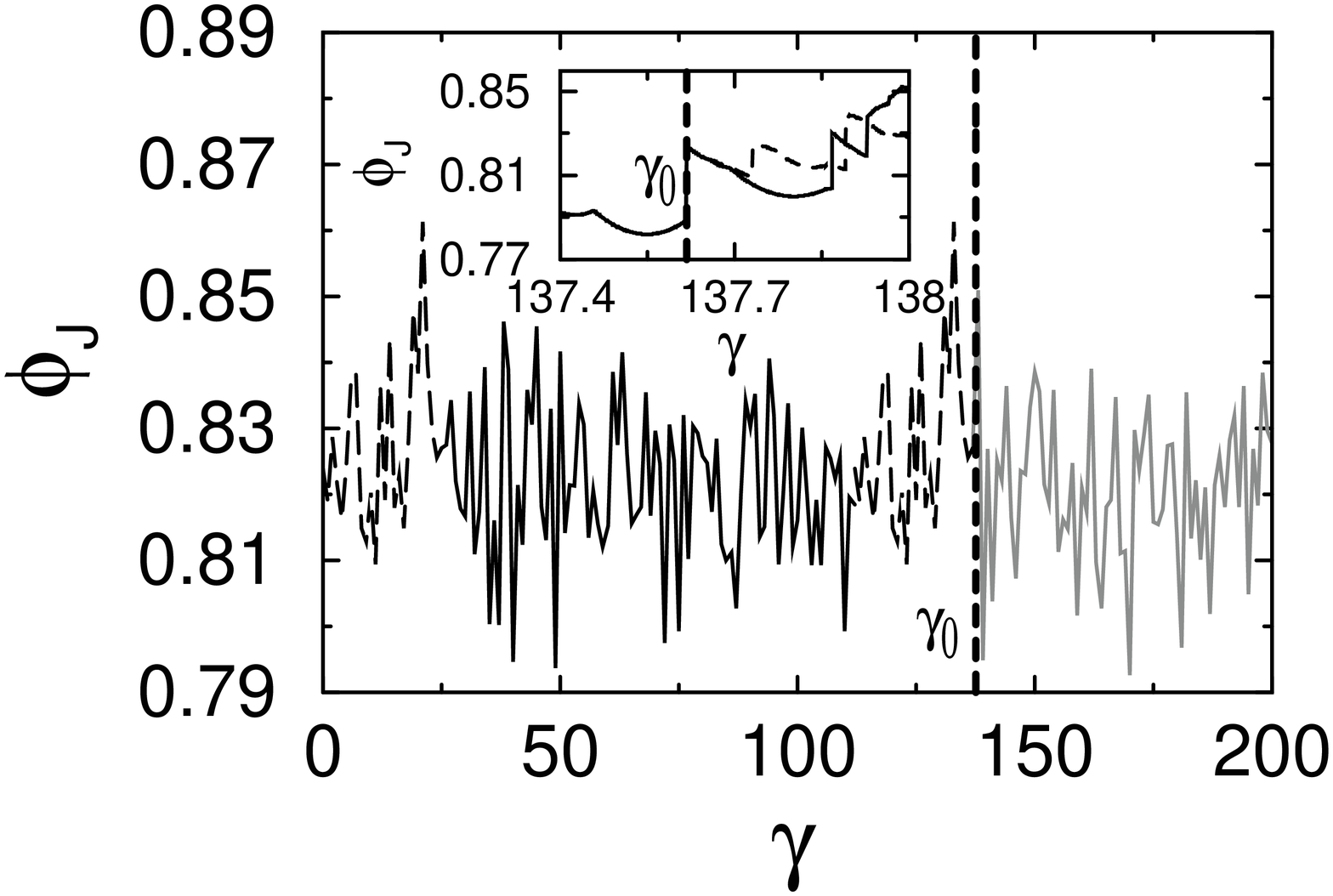}}%
\vspace{-0.1in}
\caption{\label{fig:determinism} The jammed packing fraction $\phi_J$
versus shear strain $\gamma$ during quasistatic shear flow at zero
pressure beginning from an initially unstrained MS packing for $N=20$.
After a short initial transient ($\gamma_t \sim 1$), the system locks
into an apparent periodic orbit with large period $T > 100$. The first
cycle is denoted using long-dashed and solid black lines.  However,
during the initial phase of the second cycle (long-dashed line), the
trajectory begins to deviate from that in the first cycle near
$\gamma_0=137.6$ (vertical dashed line).  The non-repeating part of
the second cycle is depicted using a solid gray line. In the inset, we
overlay cycles $1$ (dotted) and $2$ (solid) near $\gamma_0$.}
\vspace{-0.1in}
\end{figure}

\begin{figure}
\scalebox{0.3}{\includegraphics{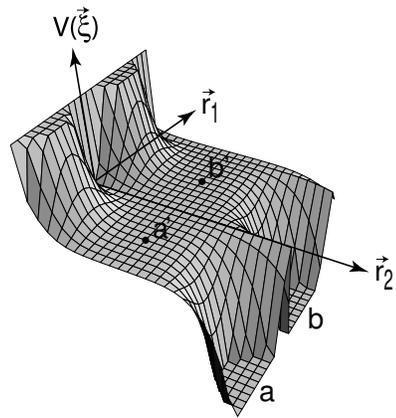}}%
\vspace{-0.1in}
\caption{\label{fig:flat} A schematic of the energy landscape
$V({\vec \xi})$, which has a flat region near local energy
minima (MS packings) $a$ and $b$.  For the quasistatic shear flow simulations,
numerical precision and specific details of the energy minimization
scheme will influence whether the system reaches point $a'$ or $b'$ in
the landscape, which in turn influences the likelihood of the 
system residing in MS packing $a$ or $b$. }
\vspace{-0.22in}
\end{figure}

\section{Trajectory splitting: Breakdown of Determinism}
\label{random splitting}

In the previous Secs.~\ref{Quasistatic evolution} and~\ref{families},
we described our results for quasistatic shear flow at zero pressure
for small systems $N \le 14$, which displayed deterministic and
contracting dynamics.  In particular, we found that when a small MS
packing undergoes a jump discontinuity at a given $\gamma$, it makes a
transition to a uniquely determined MS packing.  Thus, these systems lock
into periodic orbits with typically small periods and the number of MS packings
that are sampled at large shear strain is a small fraction of all
possible MS packings.
 
In contrast to the results found for smaller systems, we find that for
slightly larger systems ($N > 16$) there is a dramatic increase in the
period.  In addition, the deterministic behavior appears to break
down, {\it i.e.} we can not predict with unit probability the
transitions from one MS packing to another.

The breakdown of determinism is shown for $N=20$ in
Fig.~\ref{fig:determinism}, where we plot the evolution of $\phi_J$
during quasistatic shear at zero pressure.  After a short transient
strain ($\gamma_t \sim 1$), the system falls onto an apparent periodic
orbit with large period $T>100$.  However, as shown in the inset to
Fig.~\ref{fig:determinism}, during the second cycle at $\gamma_0
\approx 137.6$, the trajectory of the first and second cycles begin to
deviate.

There are several possible mechanisms for the introduction of
stochasticity and sensitivity to initial conditions in these systems,
which include bifurcations in configuration space caused by local
symmetries of the MS packings and noise (or numerical error) from the
packing-generation protocol~\cite{footnote4}.

An example of a bifurcation in the energy landscape is shown in
Fig.~\ref{fig:flat}.  This region of the landscape is extremely
flat, and thus the state of the system will depend on the numerical
precision and specific details of the energy minimization scheme.  For
example, if the energy minimization stops at point $a'$ in
Fig.~\ref{fig:flat}, it will likely move toward MS packing $a$ during
the packing-generation process, while if it stops at point $b'$, the
system may proceed to MS packing $b$.

\begin{figure}
\scalebox{0.3}{\includegraphics{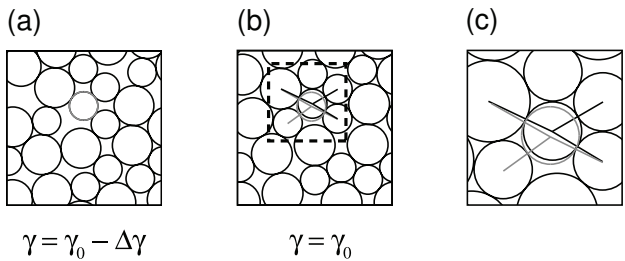}}%
\vspace{-0.5in}
\caption{\label{fig:split} Particle configurations immediately
\subfig{a} before and \subfig{b} after the trajectory splitting event
in Fig.~\ref{fig:determinism}. In \subfig{b} and \subfig{c}, the
configurations at shear strains $\gamma_0$ and $\gamma_0 - T$ are
overlayed. The only difference between the two configurations in
\subfig{b} is the position and resulting interparticle contacts
between the central and neighboring particles in the dashed box. The
particle outline and bonds are black (gray) for the configuration at
$\gamma_0$ ($\gamma_0 - T$).  Panel \subfig{c} is a magnified version of the
dashed box in \subfig{b}.}
\vspace{-0.20in}
\end{figure}

The configurational view of the trajectory splitting mechanism in
Fig.~\ref{fig:determinism} is demonstrated in Fig.~\ref{fig:split},
where we show the configuration before and after the splitting event
at $\gamma = \gamma_0$.  There are subtle differences in the position
and interparticle contacts of a single particle in the central region
of the cell at strains $\gamma_0-T$ (gray outline) and $\gamma_0$
(black outline).  This small change in the contact network, which
occurs in a flat region of the energy landscape, leads to large
differences at subsequent values of shear strain.  In this case, the cause of
the flatness in the energy landscape stems from the fact that the
directions ${\hat r}_{ij}$ for different contacting particles $j$ of a
central particle $i$ are nearly collinear.

\begin{figure}
\scalebox{0.35}{\includegraphics{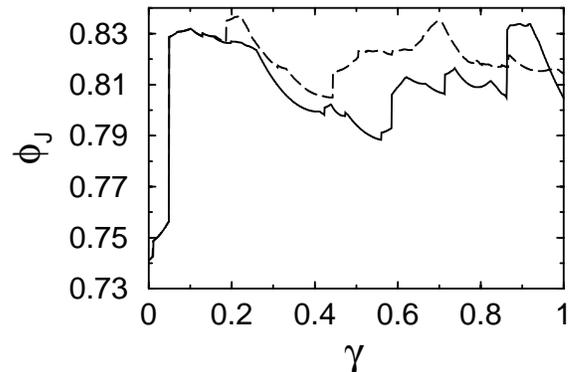}}%
\vspace{-0.18in}
\caption{\label{fig:noise} Evolution of $\phi_J$ during quasistatic
shear flow for $N=14$ for a system with (dashed line) and without (solid 
line) Gaussian noise with width $0.01$ times the small particle
diameter added to the particle positions after each strain step.}
\vspace{-0.22in}
\end{figure}

Similarly, if we add random displacements during quasistatic shear, we
can create bifurcating trajectories $\phi_J(\gamma)$.  In
Fig.~\ref{fig:noise}, we show the evolution of $\phi_J(\gamma)$ for
systems with and without added Gaussian random displacements with both
systems initialized with the same MS packing at $\gamma=0$.  We find
that the trajectory with noise deviates from the original trajectory
at $\gamma \approx 0.2$.  Thus, noise is able to compete with the
contracting mechanism to increase the fraction of dynamically sampled
MS packings.  In future studies, we will determine the fraction of MS
packings visited as a function of the noise amplitude and system size.

Rattler particles can also lead to sensitivity to initial conditions.
Small changes in the location of the rattler particle even in
MS packings that otherwise have the same network of particle contacts
can lead to large differences in subsequent contact networks if the
rattler joins the connected network in different locations.  This
effect occurs because there is no energetic incentive for rattler
particles to be located in any particular location within the
confining void region as long as it does not overlap another
particle. The influence of rattler particles on transition rates 
will also be assessed in future studies by varying the noise amplitude.

\section{Conclusions}
\label{conclusions}

In this work, we enumerated and classified the mechanically stable
packings of bidisperse frictionless disks that occur as a function of
the applied shear strain $\gamma$.  We showed that MS packings form
continuous geometrical families defined by the network of particle
contacts.

In addition, we studied the evolution of these systems during
quasistatic shear strain at zero pressure to mimic the dynamics of
slowly sheared granular media.  For small systems $N<16$, we found that the
dynamics was deterministic and strongly contracting, {\it i.e.} if the
system is initialized in an ensemble of MS packings, it will quickly
contract to at most a few MS packing families.  The strong contraction
stems from an abundance of short families, a propensity for the system
to undergo more jumps than kinks in $\phi_J(\gamma)$, the fact that
jumps only lead to increases in packing fraction, and the observation
that families at higher packing fraction attract more jumps.
 
In our studies of system sizes $N>16$, we began to see features
of large systems, including a dramatic increase in the period of the
periodic orbits and bifurcations that lead to the random splitting of
trajectories ($\phi_J$ vs. $\gamma$).  We suggest that both the
contraction and splitting mechanisms will persist in the large-system
limit, and the fraction of MS-packing geometrical families that are
visited in steady-state will
depend on ratio of the splitting and contraction rates. In large
systems, we suspect that the dynamics will focus the system onto sets
of frequent MS-packing families with similar structural and mechanical
properties, although much more work is required to quantify these
claims.

Our long-term research objective is to develop a master-equation
formalism to describe macroscopic slowly driven granular systems from
the `bottom-up' in terms of collections of small subsystems or
microstates. In this manuscript, we took a significant step forward in
this effort. We identified the types of microstates that can exist
over the full range of shear strain and studied the probabilities with
which they occur.  This information can be used as input in the
master-equation approach to calculate the contraction and splitting
rates and ultimately the steady-state distributions of macroscale MS
packings.

\begin{acknowledgments}
Financial support from NSF grant nos. CBET-0348175 (GG, JB),
DMR-0448838 (GG, CSO), and DMS-0835742 (CSO) is gratefully
acknowledged.  We also acknowledge generous amounts of CPU time from
Yale's Center for High Performance Computing.  We thank
S. Hilgenfeldt, G. Lois, M. Shattuck, and W. Zhang for insightful
comments.
\end{acknowledgments}

\appendix

\section{Numerical details}
\label{Numerical details}

In this appendix, we elaborate on technical details of the 
simulations not described in the main text.  We include specific 
numerical parameters of the packing-generation protocol and the method 
to construct the complete geometrical family map.   

\paragraph*{Packing-generation protocol}
In Sec.~\ref{Packing-generation protocol}, we outlined our procedure
to generate mechanically stable packings.  Here, we
provide some of the numerical parameters involved in the simulations.
For the energy minimization, we employ the conjugate gradient
technique~\cite{numrec}, where the particles are treated as massless.
The stopping criteria for the energy minimization ($V_{t} - V_{t-1} <
V_{\rm tol} = 10^{-16}$ and $V_t < V_{\rm min} = 10^{-16}$, where
$V_t$ is the potential energy per particle at iteration $t$) and the
target potential energy per particle of a static granular packing
($V_{\rm tol} < V < 2 V_{\rm tol}$) are the same as used in previous
studies~\cite{gao}.  For the first compression or decompression step
we use the packing-fraction increment $\Delta\phi = 10^{-4}$.  Each
time the procedure switches from expansion to contraction or vice
versa, $\Delta \phi$ is reduced by a factor of $2$.  Using the packing
generation procedure with these parameters, we are able to locate the
jamming threshold in packing fraction $\phi_J$ to within $10^{-8}$ for
each static packing over the full range of $\gamma$.  Since we implement 
an energy minimization technique with no inertia, we do not need
to alter the stopping criteria to handle rattler particles, which
possess fewer than three contacts and are not members of the force
bearing network.

\paragraph*{Geometrical Families}

To construct the complete map of geometrical families, we divided the
region $\gamma=[0,0.5]$ into small shear strain intervals
$\gamma_{s+1} - \gamma_{s} \equiv \Delta \gamma = 10^{-2}$.  For the
range of system sizes $N=4$ to $20$ studied, this choice for $\Delta \gamma$
limited the number of rearrangement events to roughly one 
per shear strain interval.  At each sampled shear strain $\gamma_s$,
we generated at least $N_t = 10^6$ MS packings using random initial
particle positions.

Two MS packings at different shear strains are considered to belong to
the same geometrical family if they possess the same set of particle
contacts.  The particle contact networks of two MS packings can be
distinguished by comparing the eigenvalues of their connectivity
matrices $C_{ij}$, where the $ij$-th element of ${\bf C}$ is $1$ if
particles $i$ and $j$ are in contact and $0$ otherwise.  Two systems
have the same contact network if all of the eigenvalues of their
connectivity matrices are the same.

\section{Dynamical matrix}
\label{matrix}

In this appendix, we calculate the elements of the dynamical matrix
(\ref{dyn_matrix}) for the repulsive linear spring potential
(\ref{spring potential}).  We employ slightly different notation for
the dynamical matrix compared to (\ref{dyn_matrix}) by separating the
spatial $\alpha,\beta=x,y$ and particle $i,j=1,\cdots,N$ indexes.  As
given in Ref.~\cite{barrat}, for pairwise, central potentials the
dynamical matrix has the following form for the off-diagonal
components $i\ne j$:
\begin{equation}
\label{offdiagonal}
M_{i\alpha,j\beta}
  =-\frac{t_{ij}}{r_{ij}}\left(\delta_{\alpha \beta}
    -{\hat r}_{ij\alpha}{\hat r}_{ij\beta}\right)
    -c_{ij} {\hat r}_{ij\alpha} {\hat r}_{ij\beta},
\end{equation}
where $\hat r_{ij\alpha}$ is the $\alpha$th component of ${\hat r}_{ij}$,
\begin{equation}
t_{ij}  \equiv \frac{{\partial V(r_{ij} )}}{{\partial r_{ij} }} = -\frac{\epsilon}{\sigma_{ij}} \left(1 - \frac{{r_{ij} }}{{\sigma _{ij} }}\right) \Theta \left(\frac{{\sigma _{ij} }}{{r_{ij} }} - 1\right),
\end{equation}
and
\begin{equation}
c_{ij}  \equiv \frac{{\partial ^2 V(r_{ij} )}}{{\partial r_{ij} ^2 }} = \frac{\epsilon}{\sigma_{ij}^2} \Theta \left(\frac{{\sigma _{ij} }}{{r_{ij} }} - 1\right).
\end{equation}
In the calculation of $t_{ij}$ and $c_{ij}$, we have ignored the
$\delta$-function contributions arising from cases when particles $i$
and $j$ are just touching with $r_{ij} = \sigma_{ij}$.  The diagonal
components ($i=j$) of the dynamical matrix are given by
\begin{equation}
M_{i\alpha,i\beta} = - \sum_{j=1,i\ne j}^N M_{i\alpha,j\beta}.
\end{equation}
The shear-periodic boundary conditions only affect the definition of 
the separation $r_{ij}$ for particles near the edges of the simulation 
cell.

\section{Polarizations of MS packings}
\label{Appendix: polarizations}

In this appendix, we provide details for generating different 
polarizations in simulations and determining which polarizations 
are distinct.  
  
\begin{figure}
\includegraphics[width=0.4\textwidth]{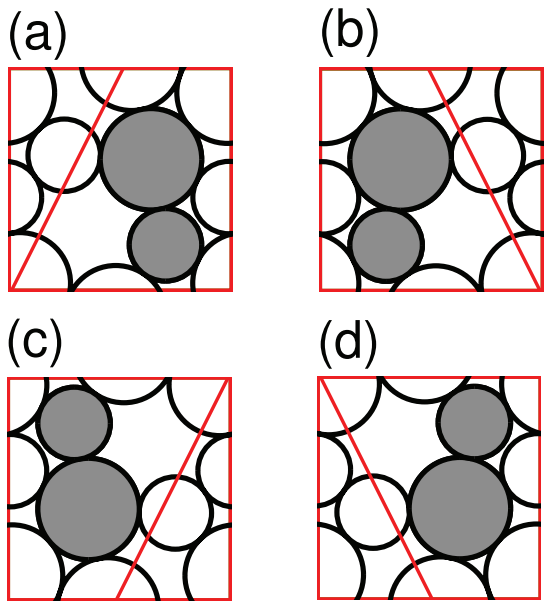}
\vspace{-0.1in}
\caption{\label{fig:2polar} \subfig{a} Typical MS packing at 
half-integer shear strains.  We apply the transformations that are
consistent with shear-periodic boundary conditions---the
roto-inversion transformations ${\bf R}_3$, ${\bf R}_6$, and ${\bf
R}_8$ in Table~\ref{tab:table_noshear}---to the configuration in
\subfig{a} to generate the configurations in
\subfig{b}-\subfig{d}. However, \subfig{a} and \subfig{b} and \subfig{c}
and \subfig{d} are related by the shear symmetry operation (rotations
by $\pi$, ${\bf R}_3$), and thus there are only two distinct
polarizations at half integer strains.  }
\end{figure}

\begin{table}
\caption{\label{tab:table_noshear} The eight roto-inversion matrices
${\bf R}_i$ in 2D that when applied to a given MS packing at integer
values of shear strain generate the eight equivalent polarizations
with identical eigenvalue lists of the dynamical matrix.}
\begin{ruledtabular}
\begin{tabular}{rrrrcc}

(1) $\left( {\begin{array}{*{20}c}
      1 & 0  \\
      0 & 1  \\
    \end{array}} \right)$& 
(2) $\left( {\begin{array}{*{20}c}
      0 & 1  \\
      -1 & 0  \\
    \end{array}} \right)$& 
(3) $\left( {\begin{array}{*{20}c}
      -1 & 0  \\
      0 & -1  \\
    \end{array}} \right)$& 
(4) $\left( {\begin{array}{*{20}c}
      0 & -1  \\
      1 & 0  \\
    \end{array}} \right)$\\

(5) $\left( {\begin{array}{*{20}c}
      0 & 1  \\
      1 & 0  \\
    \end{array}} \right)$& 
(6) $\left( {\begin{array}{*{20}c}
      -1 & 0  \\
      0 & 1  \\
    \end{array}} \right)$& 
(7) $\left( {\begin{array}{*{20}c}
      0 & -1  \\
      -1 & 0  \\
    \end{array}} \right)$& 
(8) $\left( {\begin{array}{*{20}c}
      1 & 0  \\
      0 & -1  \\
    \end{array}} \right)$\\

\end{tabular}
\end{ruledtabular}
\end{table}

We will first describe the symmetries that MS packings possess under
shear periodic boundary conditions in undeformed cells, since these
symmetries affect the number and types of MS packings that can be
obtained during shear.  At integer shear strains, there are eight
`polarizations' all of which have the same list of eigenvalues of the
dynamical matrix, but different eigenvectors.  A given MS packing at
integer shear strain (panel \subfig{1}) and its equivalent
polarizations (panels \subfig{2}-\subfig{8}) are shown in
Fig.~\ref{fig:polarization_noshear}.  Each polarization with
coordinates ${\vec \xi}_i$ in panel \subfig{i} in
Fig.~\ref{fig:polarization_noshear} can be obtained from the original
MS packing ${\vec \xi}_1$ in panel \subfig{1} using
\begin{equation}
\label{rotation}
{\vec \xi}_i = {\bf R}_i {\vec \xi}_1, 
\end{equation} 
where the eight roto-inversion matrices ${\bf R}_i$ with ${\rm det} {\bf
R}_i = \pm 1$ in 2D are given in Table~\ref{tab:table_noshear}.  In
isotropically compressed systems these polarizations occur with equal
probability.  However, in systems subjected to shear strain, these
polarizations will occur with different probabilities.

2D systems subjected to simple planar shear flow possess a discrete
rotational symmetry; {\it i.e.} the system is unchanged when it is rotated
by $\pi$ about an axis coming out of the page ({\it i.e.} apply ${\bf
R}_3$ to a given MS packing) as shown in the bottom of
Fig.~\ref{fig:polarization_noshear}.  In panels
\subfig{1}-\subfig{8} in Fig.~\ref{fig:polarization_noshear}, 
we see that polarizations $1$ and $5$, $2$ and $6$, $3$ and $7$, and
$4$ and $8$ are related by rotations by $\pi$, and therefore will
behave the same under simple planar shear.  Thus, only four distinct
polarizations at integer shear strains remain.

If the accumulated shear strain is half-integer, MS packings will have
at most only two distinct polarizations. Only four roto-inversion
transformations are consistent with shear-periodic boundary
conditions, ${\bf R}_3$, ${\bf R}_6$, and ${\bf R}_8$ in
Table~\ref{tab:table_noshear}. These transformations have been applied
to the configuration \subfig{a} in Fig.~\ref{fig:2polar} to
generate the configurations
\subfig{b}-\subfig{d}. However, \subfig{a} and \subfig{b} and \subfig{c}
and \subfig{d} are related by the shear symmetry operation (rotations
by $\pi$, ${\bf R}_3$), and thus only two distinct polarizations
remain at half integer strains.

In general, there is only one polarization for all shear strains other
than integer and half-integer.  Moreover, the number of polarizations
at integer (half-integer) strains can be smaller than four (two) if
particular MS packings possess additional symmetries.  

In simulations, we distinguish the polarizations of two MS packings by
applying all possible roto-inversion transformations configurations
consistent with the shear-periodic boundary conditions to a given MS
packing. We then compare the eigenvector (corresponding to the
smallest nonzero eigenvalue) of the second configuration to that for
all of the roto-inverted configurations of the first and look for a 
match.


\begin{references}
\bibitem{footnote_1} In our
simulations we use the criterion $|m_i| > m_{\rm min}$ for nonzero
eigenvalues, where $m_{\rm min} = 10^{-6}$ is the noise threshold for our
eigenvalue calculations.  Also, eigenvalues are considered to be
equal if they differ by less than the noise threshold $m_{\rm min}$.
\bibitem{foot} We estimate that only $\sim 60\%$ of the total number
of distinct MS packings $N_s^p(\gamma)$ were found for $N=10$.
\bibitem{footnote2} We have also calculated the family length
distribution using the eigenvalues of the connectivity matrix to
determine the strains at which geometrical families begin and end,
which is identical to the length distribution shown in
Fig.~\ref{fig:binned_family_length} (a). (See Appendix~\ref{Numerical
details}.)
\bibitem{footnote4}
It is also possible that plotting $\phi_J$ does not
adequately distinguish one MS packing from another; however, we have
verified that the sequence of eigenvalue lists and polarizations at
the beginning of the second cycle match those at the beginning of the
first.
\end{references}
\end{document}